\documentclass[english,aps,reprint,showpacs,titlepage,longbibliography,floatfix,prstper]{revtex4-2}   

\usepackage[T1]{fontenc}	
\usepackage[latin9]{inputenc}	
\usepackage{geometry}		
\usepackage{graphicx}
\usepackage[above,below]{placeins}	
\usepackage{times}
\usepackage{xfrac}
\usepackage{booktabs}
\usepackage{hyperref}  
\hypersetup{colorlinks=true,urlcolor=blue,citecolor=blue,linkcolor=blue}   
\usepackage{enumitem}          
\setlist{nosep}                 
\usepackage{xcolor}
\usepackage{multirow}
\usepackage{amssymb}
\usepackage{array}
\usepackage{makecell}
\usepackage{amsmath}
\newcolumntype{K}[1]{>{\centering\arraybackslash}p{#1}}

\begin{document}

\begin{titlepage}

    \title{Landscape of Quantum Information Science and Engineering Education: From Physics Foundations to Interdisciplinary Frontiers}

  \author{A.R. Pi\~na\textsuperscript{1}} \email{andi.pina@rit.edu} \altaffiliation{they/them/theirs}
  \author{Shams El-Adawy\textsuperscript{2,3}} 
  \author{Mike Verostek\textsuperscript{1}}
    \author{Brett T. Boyle\textsuperscript{1}}
    \author{Mateo Cacheiro\textsuperscript{1}}
    \author{Matt Lawler\textsuperscript{1}}
    \author{Namitha Pradeep\textsuperscript{1}}
    \author{Ella Watts\textsuperscript{1}}
    \author{Colin G. West\textsuperscript{3}}
    \author{H. J. Lewandowski\textsuperscript{2,3}}
    \author{Benjamin M. Zwickl\textsuperscript{1}}

  \affiliation{\textsuperscript{1}School of Physics and Astronomy, Rochester Institute of Technology, Rochester, NY 14623, USA}
  \affiliation{\textsuperscript{2}JILA, National Institute of Standards and Technology and the University of Colorado, Boulder, Colorado 80309, USA}
  \affiliation{\textsuperscript{3}Department of Physics, University of Colorado, Boulder, Colorado 80309, USA }


 \begin{abstract}
Quantum Information Science and Engineering (QISE) is rapidly gaining interest from those within many disciplines and higher education needs to adapt to the changing landscape.
Although QISE education still has a strong presence and roots in physics, the field is becoming increasingly interdisciplinary. 
There is a need to understand the presence of QISE instruction and quantum-related instruction across all disciplines in order to figure out where QISE education is already happening and where it could be expanded.  
Although there is recent work that characterizes introductory QISE courses, there is no holistic picture of the landscape of QISE and quantum-related education in the United States. 
To understand how QISE education is evolving, we analyzed course catalogs from 1,456 U.S. institutions. 
We found 61 institutions offering QISE degree programs, mostly at PhD-granting schools, with interdisciplinary programs being the most common. 
Physics, electrical and computer engineering, and computer science are the primary contributors to these programs.
Across all institutions, we identified over 8,000 courses mentioning 'quantum,' mainly in physics and chemistry departments, but about one-third of institutions in our study had none. 
We also found over 500 dedicated QISE courses, concentrated in PhD-granting institutions, primarily in physics, electrical and computer engineering, and computer science.
Physics leads in offering both general quantum-related ($\sim$4,700) and QISE-specific ($\sim$200) courses. 
While quantum knowledge is often a prerequisite for QISE, there are efforts to make QISE education more accessible to students with less background in quantum mechanics.
Across multiple disciplines, we also see that QISE topics are being introduced in courses not fully dedicated to QISE, which may be a productive strategy for increasing students' access to QISE education. 
Our dataset and analysis provide the most comprehensive overview to date of quantum education across US higher education. 
To ensure broad access, all data are publicly available and downloadable at quantumlandscape.streamlit.app. 
We hope these findings will support and guide future efforts in curriculum design, workforce development, and education policy across the quantum ecosystem.

  \end{abstract}

  \maketitle
\end{titlepage}

\section{Introduction}
\subsection{Motivation}

Quantum Information Science and Engineering (QISE) is a rapidly growing field in both academia and industry\footnote{A note on terminology: There are multiple ways people refer to quantum information. 
QIS is Quantum Information Science.
Some reports add a ``T'' which stands for Technology. Throughout the paper, we will use QISE to encompass all of these instances, aside from direct quotations.}. 
Over the past seven years, quantum technology has emerged as a federal priority, which has increased the investment in QISE research and education. 
In 2018, the National Quantum Initiative (NQI) Act authorized funding and called for multiple federal agencies to support research and development centers for quantum technology research \cite{NQI}. 
The NQI Act  emphasized primarily research and technology development, while also recognizing the need for a QISE workforce `pipeline,' but with few details on how that should be accomplished. 

In 2022, building off of the NQI Act, the Quantum Information Science and Technology Workforce Development National Strategic Plan discussed opportunities for academia to contribute to this development.
The plan states,  ``Institutions of higher education can expand QIST  
courses and programs to increase opportunities for future workers and to build proficiencies that connect various specializations with QIST expertise.'' \cite{NSP}
Furthermore, the 2022 CHIPS and Science Act makes a call to ``assess the state of quantum information science education and skills training at all educational levels and identify gaps in meeting current and future workforce needs,'' recognizing that a crucial step in the development of educational materials is developing an understanding of what already exists \cite{CHIPS}.

In addition to this federal priority on quantum education and workforce development, physics educators and education researchers have expanded their efforts in this area as well.
The physics education research community has been increasingly focused different aspects of quantum-related education such as difficulties \cite{Singh2008,Singh2013, Singh2014a, Marshman2015, Marshman2015a, Singh2015a},  conceptual understanding \cite{Gire2008, Gire2011,Gire2012, Passante2015a, Passante2015b, Emigh2015,Emigh2016a, Hoehn2018, Passante2020,Corsiglia2023, Riihiluoma2024}, understanding of notations \cite{Gire2015a, Schermerhorn2019, Wawro2020}, mathematical sensemaking \cite{Dreyfus2017, Pina2023, Pina2024},  understanding of visualizations \cite{Kohnle2017, Passante2019,Ahmed_2022}, development of instructional materials \cite{Singh2016a, Emigh2018, Schermerhorn2022}, and quantum experiments \cite{Borish2023_qm_effects_in_exp, Borish2024_student_reasoning_during_experiments, Oliver_qforge_2025}.
QISE-specific education efforts in the field have included development of instructional materials \cite{DeVore2015}, studies on workforce development \cite{Fox2020_workforce, hughes2022, hasanovic2022,Oliver_qforge_2025}, and landscape studies \cite{Cervantes2021, Meyer2024}.

Although there is ongoing work on the characterization of the content \cite{meyer_intro} and availability \cite{Cervantes2021, Meyer2024} of introductory quantum and QISE courses, there is no holistic picture of quantum-related education in the United States (US). 
The needs of the QISE industry are also being actively studied in both US \cite{Fox2020_workforce, hughes2022, hasanovic2022} and European \cite{greinart2023_workforce, greinert_advancing_2024} contexts in order to inform the further development of QISE education.
A deeper understanding of the needs of the QISE industry is crucial to preparing students for entering the workforce, however a fuller picture of existing QISE education is required for educators to understand where to implement curricular changes. 

The work presented here builds on previous work \cite{Cervantes2021, meyer_intro, Meyer2024} on the characterization of the landscape of QISE courses, programs, and other QISE educational activities in higher education, which will be discussed in more detail in the next section. 
In addition to updating the findings of previous studies to reflect the changing landscape, this project also uses the largest sample of institutions to date and should provide a reasonably  comprehensive picture of the educational landscape in the US. 
Previous studies have also been focused strictly on QISE courses, whereas the data presented herein are on all courses instructing in quantum topics. 

Although QISE is increasingly interdisciplinary, physics is the historical home of both quantum and quantum information.
In both this study and prior studies, it has been shown that physics is one of the largest, if not the largest, contributors to QISE coursework, which has motivated us to examine specifically contributions of physics departments  to QISE. 
Our goal is to provide information about current quantum-related and QISE courses and QISE-focused programs to support the work of various stakeholders.

For curriculum developers, this information should be useful in determining what other institutions/instructors are focusing on in their courses and programs. 
Access to information on different programs across the nation can aid leaders in the development and evolution of QISE programs at their institutions.
As more institutions develop QISE programs, these data can also help an institution find what sets it apart from existing programs. 
For instructors, this can be a resource for determining which topics are of interest to a wide variety of disciplines. 
These data also potentially provide significant aid to students seeking educational opportunities through self-exploration, or to academic advisors and faculty members guiding students' educational and career decisions. 
This work is not about assessing any particular course or program, but rather provides national context to an increasingly important and rapidly changing field.

The data presented here allow us to address the following research questions:
\begin{itemize}
    \item[\textbf{RQ1:}] What are the characteristics of QISE programs in higher education across the US?  
    \item [\textbf{RQ2}] What is the distribution of courses that include quantum and/or QISE across different disciplines, modalities, and institution types? 
    \item[\textbf{RQ3:}] Within physics departments, in what ways does the distribution vary by level, modality, and type of course for quantum and QISE topics? 
\end{itemize}
The two primary characteristics that we consider with regard to institution type are the Minority Serving Institution (MSI) status and Carnegie classification of Institutions of Higher Education. 
Course level is differentiated by graduate versus undergraduate instruction, as well as prerequisites for QISE courses. 
Modality refers to whether a course is primarily lecture, lab, or a mix of the two. 
\footnote{Note that we do not have a means of differentiating between different instructional methods (e.g., traditional lecture, flipped classroom, etc.).} 

\subsection{Background}
Two recent studies have investigated QISE courses in the US. 
\citet{Cervantes2021} searched 2019~\textendash~2020 course catalogs for QISE courses at 305 US institutions in order to examine the availability of courses at different levels and how that varied based on institutional characteristics, such as Carnegie classification and status as a MSI. 
Within that sample, they identified 74 institutions with QISE courses. 
Of those institutions, a significant majority (64) were PhD granting institutions. 
Regarding MSI status, the study found QISE courses at Hispanic Serving Institutions and Asian American Native American Pacific Islander Serving Institutions, however,they did not find any courses at any other type of MSI.
Their findings suggested that there may be a disparity in access to QISE coursework based on factors such as MSI status or institutional resources.
There were two primary limitations of this work, both of which were explicitly addressed by the authors. 
The first is that this is a relatively small sample given the total number ($\sim4100$) of post-secondary institutions in the US and the study included a small fraction of the MSIs (56 out of 711). 
The second is that the study addressed only QISE-specific courses, which does not capture the extent to which QISE instruction may be occurring in courses not dedicated to QISE, such as a traditional quantum mechanics course with a unit on quantum computing and cryptography. 

 \citet{Meyer2024} directly followed up on the work of \citet{Cervantes2021} by expanding the sample to courses offered in the 2019~\textendash~2020 academic year at 456 US institutions.
 Their selection criteria for these institutions were centered around degrees granted in physics, computer science, and electrical/electrical and computer engineering programs that either granted the most degrees or met a minimum threshold depending on degree level.
However, much like \citet{Cervantes2021}, \citet{Meyer2024} looked exclusively for QISE-specific courses. 
With these data, they also performed a multiple linear regression in addition to some descriptive statistics. 
Their regression examined Carnegie classification, funding, religious affiliation, MSI status, percentage of students receiving Pell Grants, and urbanization index of the institution as independent variables with the dependent variables being availability of QISE courses and programs. 

The regression analysis confirmed the findings of \citet{Cervantes2021} that QISE courses were found disproportionately at research-intensive institutions (commonly known as R1/Doctoral granting very-high research activity and R2/Doctoral-granting high research activity). 
Research intensive institutions were shown to be approximately nine times more likely to offer to QISE courses than non-research intensive institutions \citet{Meyer2024}. 
There was also a large disparity in access found between public and private institutions, with public institutions being significantly less likely to offer QISE courses; for public institutions in more rural areas, this effect was even greater. 
The independent variable related to the percent of students receiving Pell Grants was used a proxy for institutional resources and showed that access to QISE coursework decreased as the percent increased.
When controlling for the other independent variables in the model, no significant association was found with an institution's MSI status. 
These findings suggest that the most significant factor in determining the likelihood of an institution to offer QISE courses is access to resources, such as faculty expertise and funding.  

Some efforts have been made to characterize a subset of the landscape of QISE education beyond just the US. 
\citet{goorney_2025} conducted a global search for QISE master's programs and identified 86 institutions with such programs. 
The majority of the institutions identified are located in the EU (41) or US (17).
Unlike the work of \citet{Meyer2024}, which reported on QISE education at all levels, \citet{goorney_2025} reported exclusively on the graduate level to explore whether master's programs can provide an easy transition to industry. 
Some of the findings of \citet{goorney_2025} suggest that there may be increasing opportunities for those with QISE master's degrees to find positions in industry.

Two groups have been working on relevant course content analysis. 
The first group, \citet{meyer_intro}, looked into the content of introductory QISE courses. 
They surveyed instructors of 63 introductory QISE courses at US institutions listed in physics (26 courses), computer science (13 courses), and electrical and computer engineering (11 courses), as well as 11 courses cross-listed among different combinations of those disciplines. 
This survey identified a collection of commonly taught QISE topics that are being used as the basis for a concept inventory of introductory quantum information science concepts.
The focus of \citet{meyer_intro} on QISE courses  does not account for the possibility that QISE topics may be covered in other courses not solely dedicated to QISE. 

An example of QISE topics in non-dedicated courses can be found in the work of the second group, \citet{buzzell_modern_2024, buzzell_quantum_2024}, who have been investigating topics taught in modern physics and quantum mechanics courses in the US. 
They have leveraged large language models to perform large-scale syllabus analysis for 167 modern physics courses and 56 quantum mechanics courses. 
Within modern physics courses, quantum physics is the most commonly covered topic.
This analysis has revealed that quantum mechanics is the most common topic covered in modern physics courses. 
In half of the quantum mechanics courses (28/56), the syllabi included QISE topics such as quantum computing, quantum teleportation, and quantum cryptography.

The work presented herein expands on these other studies by examining more institutions, and searching a broader selection of courses, than any previous study to date. 
In addition to looking for QISE courses, we have searched for all instances of `quantum' in course titles and descriptions, as well as program names in a sample of 1,456 institutions.. 
This analysis allowed us to examine the growing landscape of QISE courses reported by \citet{Cervantes2021} and \citet{Meyer2024} and present a broader picture of institutions in the US. 
Considering all instances of quantum-related instruction, in addition to QISE, allows us to identify existing courses where QISE could be added as an application or unit within the course. 
As mentioned in \citet{buzzell_quantum_2024}, there is already a precedent for QISE topics appearing in non-dedicated courses (e.g. quantum mechanics). 
We also examined the prerequisites for different QISE courses and entrance requirements for different QIS programs. 
These data highlight the different emergent philosophies about the types of information that are necessary for a student to learn QISE topics. 

\section{Methods\label{sec:methods}}

\subsection{Institutional Selection Criteria}
The first step in carrying out the project was to decide which institutions to include in our sample. 
There are more than 6,000 institutions of higher education across the United States.
Approximately 4,100 of those institutions grant undergraduate and graduate degrees across a range of subjects, while the remaining institutions focus primarily on specific occupational training or religious instruction. 
Although examining all 4,100 institutions would be the most thorough approach, it is not practical. 
We therefore needed to prioritize different types of institutions. 

Previous research has shown that the majority of QISE instruction is offered at research intensive institutions \cite{Meyer2024, Cervantes2021}; therefore, we included all high and very-high research activity institutions as defined by the Carnegie classification . 
The research focus of these institutions attracts QISE experts who start and manage QISE research and educational programs. 
We aimed to create a geographically diverse sample of institutions, ensuring representation from every state in the US, to provide a comprehensive snapshot of the current educational landscape.
Additionally, community colleges, two-year colleges, and four-year colleges have been underrepresented in prior landscape studies, so we wanted to ensure a large number of these types of institutions were included. 
We specifically focused on institutions that produce graduates in fields that most commonly lead to careers in QISE, such as physics, engineering, and computer science. 
Finally, legislation and strategic planning documents that motivate some of this work \cite{NQI, CHIPS, NSP} call for efforts to ensure the broadest, skilled QISE workforce, and MSIs \cite{nguyen2023} play a central role in this effort. 
MSIs can be categorized as either mission-based, such as Historically Black Colleges and Universities (HBCUs) and Tribal Colleges and Universities (TCUs) or enrollment-based (all other MSIs). 
Mission-based MSIs explicitly set out to serve a specific student population, while enrollment-based MSIs meet criteria set by congress in terms of what percentage of their students are of a given demographic. \cite{nguyen2023}

These various considerations resulted in five inclusion criteria for institutions. 
Any institution that met at least one of these criteria was added to the overall sample.

 \begin{enumerate}
    \item Institutions with a Carnegie classification of either ``Very High Research Activity'' or  ``High Research Activity''
    \item Top 10 STEM bachelor-producing institutions by state
    \item Top 5 STEM associate-producing institutions by state
    \item Institutions with an program accredited by the Accreditation Board for Engineering and Technology (ABET) in computer science or engineering
    \item All MSIs
\end{enumerate}

We identified institutions that met criteria 1~\textendash~3 through using the National Center for Education Statistics Integrated Postsecondary Education Data System (IPEDS) \citet{IPEDS}. 
IPEDS contains institution names, locations, Carnegie classifications, and breakdowns of discipline and types of degrees awarded, in addition to other information. 
In criteria 2 and 3, STEM refers to a collection of majors that are primary contributors to QISE, including physics, electrical and computer engineering\footnote{We recognize that at some institutions electrical engineering and computer engineering are two separate departments. To remain consistent in our handling of the data, any electrical engineering, computer engineering, and electrical and computer engineering are all included in the same category.}, and computer science \cite{meyer_intro}. 
The 21 distinct Carnegie classifications were narrowed down to just four categories based on either the highest or largest number of degrees granted for ease of interpretability.
The full details of the Carnegie classifications system we used can be found in Appendix~\ref{carnegie_simp_sec}.

ABET accreditation was identified through a publicly accessible ABET-managed database with information on all institutions with current and past ABET accreditation \cite{abet}. 
MSIs were identified through NASA's Minority Serving Institutions Exchange, which includes a comprehensive listings of MSIs and their MSI classifications (i.e., what populations the institution is serving) \cite{nasa_msi}. 
These institutions include Historically Black Colleges and Universities (HBCUs), Hispanic Serving Institutions(HSIs), Asian American and Pacific Islander Serving Institutions(AAPISIs), Tribal Colleges and Universities(TCUs), Alaskan Native and Native Hawaiian Serving Institutions(ANNHSIs), and Non-Tribal Native American Serving Institutions (NASNTIs). 

Institutional information from each of these individual sources was standardized, aggregated, and cross-linked using Python code, resulting in a sample of $N = 1456$ institutions. 
A brief overview of the number of institutions in our sample based on Carnegie classification and MSI status is found in Table~\ref{tab:brief_sample_breakdown}.
A more detailed breakdown of the number of institutions meeting different combinations of the five selection criteria can be found in Appendix~\ref{sample_characteristics}.
The sample includes a larger number of two-year institutions than one might expect from the selection criteria, but this is due to the large number of two-year institutions which are MSIs. 
Similarly, based on criterion 2, one might expect the number of four-year institutions to come close to 500 (10 for each of the 50 states) but criterion 2 refers to a quantity of bachelor's degrees granted as opposed to an institution type. 
There are in fact $\sim500$ institutions meeting criterion 2, but many of them are either PhD or master's granting institutions. 
In total, institutions in this sample represent 86\% of the physics bachelors degrees awarded in the US in 2022 (the most recent IPEDS data available at the time of institution selection (June 2024), 93\% of those in computer science, and 98\% of those in engineering generally. 
Given our focus on institutions with a strong STEM presence, and some of the low percentages in the rightmost column of Table~\ref{tab:brief_sample_breakdown}, it is likely that the data in this work represent a best case scenario for course availability across different institutions.

\begin{table}[htb]
\caption{Number of institutions in our study based on Carnegie classification and MSI status. The last column indicates the percentage of institutions in our sample relative to all US institutions with that Carnegie classification. For example, the 35\% (bottom-right) means that this study contains 35\% of all degree granting institutions in the US (roughly 4,000 in total).}
\begin{tabular}{c c c c}
\textbf{Carnegie}                 & \textbf{MSI - No} & \textbf{MSI - Yes} & \textbf{\% of Total in US}  \\
\hline
\hline
\textbf{Two-Year}                 & 182               & 328                & 32\%                                \\
\textbf{Four-Year}                & 122               & 147                & 10\%                                \\
\textbf{Master's granting}                  & 207               & 132                & 20\%                                \\
\textbf{PhD granting}             & 273               & 106                & 82\%                                \\
\hline
\textbf{\% of Total in US} & 24\%              & 100\%              & \textbf{35\%}
\end{tabular}
\label{tab:brief_sample_breakdown}
\end{table}

\subsection{Collection of Course and Program Data}
Data were collected by a team consisting of two postdocs, a staff researcher, one graduate student, and four undergraduate researchers. 
The data collection effort stretched over approximately 13 weeks beginning in June 2024 and required over 600 person-hours to complete. 
The process of investigating an institution started with using an internet search engine (e.g., Google) to locate the institution's course catalogs and program catalogs. 

We searched program catalogs for the word `quantum' and recorded any program with a name containing the word `quantum' for further review as a QISE program. 
Programs were defined to include associate's, bachelor's, master's, and PhD degrees, certificates and minors, and concentrations, focuses, or tracks within majors. 
After searching a program catalog, the team returned to an internet search engine and input institution names in combination with key words (quantum computing, quantum information, quantum technology) to ensure we did not miss programs that may not appear in catalog searches. 
Course catalogs were then searched for any courses whose title or description included the word `quantum'. 
The title, course number, description, prerequisites, department(s), and level were recorded for courses that met this condition. 
Due to the lack of standardization of course catalogs, prerequisite information was not always available. 

All data were input into a Qualtrics survey to ensure uniformity and subsequently integrated with the institutional data described above. 
The aggregate course, program, and institutional data form the basis for the analysis described herein.

\subsection{Course Categorization}
In order to better understand the broader structure of all the recorded courses, we organized them into three primary groups. 
The first was just the set of all recorded courses; this will be referred to as all courses with `quantum' throughout the paper. 
The second group, labeled `courses with QISE topics,' included courses that had at least one keyword associated with QISE topics (see Appendix \ref{keywords}). 
We also performed a more fine grained classification of all recorded courses into thematic categories, one of which was the third group of dedicated QISE courses.
The first step in this categorization involved performing literal string searches on course titles to group similar courses (e.g., quantum mechanics, modern physics, quantum computing, quantum algorithms, etc.). 
This approach successfully categorized approximately 90\% of courses based on keyword matches in their titles.
The remaining courses were manually categorized by reviewing both the course title and description. 
Each course description was then checked to ensure that it belong to its respective category. 
This fine-grained classification resulted in $\sim140$ different categories for the whole set of $\sim8000$ courses. 
The individual categories were then hierarchically organized by discipline. 
This needed to be done for different disciplines independently due to the fact that disciplines tend to have unique courses that do not align with courses from other disciplines.
One exception to this was the category of QISE courses, which across disciplines had quite similar courses. 
The physics hierarchy is summarized in Fig.~\ref{tab:course_categorization}. 
Within the physics hierarchy there are a few collections of courses we would like to clarify. 
The `QISE adjacent' category includes courses that could easily be included as physics electives, however, these courses cover material that is directly applicable to QISE content, and we therefore left them separate to determine the extent to which they are addressing QISE content. 
Within the quantum computing and information category, there are subcategories with small distinctions. 
Quantum computing, information, and error correction courses all cover quantum computing to some extent; the difference between courses in these categorizes lies in their focus. 
Quantum computing courses exclusively discuss quantum computing, whereas quantum information courses mention other applications of quantum technology, such as quantum communication. 
Courses on quantum error correction discuss quantum computing architectures, but focus on primarily quantum error correction within those platforms. 
The implementation of quantum technology category includes courses that focus on either the hardware necessary for the realization of different quantum technologies or methods of controlling quantum systems in electronics. 
Note that the second to last category in the hierarchy is `Chemistry.'
This is a result of either cross-listed courses or joint departments (e.g., a department of physics and chemistry).
\begin{figure*}[htb]
\renewcommand{\arraystretch}{2}
    \centering
\begin{tabular}{|c|l|m{4cm}m{4cm}|}
\hline
\multicolumn{1}{|l|}{{\multirow{3}{*}{Physics}}} 

& \multicolumn{1}{l|}{
   {\multirow{1}{*}{Core Physics}}
}&
\textbullet~Introductory Physics

\textbullet~Modern Physics

\textbullet~Stat. Mech.  \& Thermo.

\textbullet~Electricity and Magnetism
 & 

\textbullet~Quantum Mechanics

   \textbullet~Math Methods
   
  \textbullet~Classical Mechanics

 \\
\cline{2-4} 

& \multicolumn{1}{l|}{{\multirow{1}{*}{Physics Electives}}}     

&

\textbullet~Optics

\textbullet~Biophysics

\textbullet~Nano-

\textbullet~Space Science

\textbullet~Nuclear and Particle

\textbullet~Atomic and Molecular
& 

\textbullet~Group/Field Theories

\textbullet~QED

\textbullet~Information Theory

\textbullet~Materials

\textbullet~Computational

 \\ 

\cline{2-4} 
\multicolumn{1}{|c|}{} & 

\multicolumn{1}{l|}{QISE Adjacent} &

\textbullet~AMO

\textbullet~Laser Physics
&

\textbullet~Quantum Optics  

\\ 
\hline

\multicolumn{1}{|l|}{\multirow{3}{*}{QISE}}    

&
\makecell[l]{Quantum Computing \\ and Information}
&
\textbullet~Quantum Computing

\textbullet~Quantum Entanglement

\textbullet~Quantum Algorithms
&
\textbullet~Quantum Error Correction

\textbullet~Quantum Information

\\ 

\cline{2-4} 

& \makecell[l]{Implementation of \\ Quantum Technology} & 

\textbullet~Quantum Electronics

\textbullet~Quantum Hardware             
& 

\textbullet~Superconducting Circuits 

\\ 
\cline{2-4} 

& Other QISE
& 
\textbullet~Quantum Software 

\textbullet~Quantum Simulation

\textbullet~Quantum Cryptography
&

\textbullet~Quantum Engineering

\textbullet~Research/Internship

\\ 
 
 \hline
 \multicolumn{1}{|l|}{{\multirow{2}{*}{Other}}} 
 
 &
\multicolumn{1}{l|}{Chemistry}
& 
\textbullet~Physical Chemistry

\textbullet~Inorganic Chemistry
&
\textbullet~Analytical Chemistry

\textbullet~General Chemistry
\\ \cline{2-4}
&
\multicolumn{3}{l|}{Miscellaneous}\\ 
\hline
\end{tabular}
 \caption{Hierarchical categorization of courses listed in physics departments. Individual courses were sorted into the categories in the rightmost column. The categories were then grouped as shown.}
    \label{tab:course_categorization}
\end{figure*}

\subsection{Limitations}
The methods discussed above present some inherent limitations. 
One of the most salient limitations is our use of course catalogs and descriptions. 
Course catalogs at some institutions are  updated infrequently. 
Course descriptions can also be vague; this is sometimes intentional to provide instructors with flexibility, but could also be a result of little attention being paid to them. 
These factors could lead to inaccurate data about courses and programs. 
Since QISE emerged as a field relatively recently, we believe this is likely to result in under-reporting of courses that include QISE topics, where course changes have yet to make it into publicly available course catalogs. 
This may be particularly true for  special topics courses, which are often listed with just a general description that may not address the content of the course at all. 

Due to the size of the data set, binning has been necessary at different stages of the project, but could result in a loss of some data resolution. 
The size of the data set required some use of computational tools in analysis. 
Although our use of literal string searches allowed for a great degree of uniformity in sorting courses and detecting specific topics mentioned in course descriptions, they cannot account for all potential misspellings or abbreviations that could be present in any text. 
There are packages for inexact matching of strings that are meant to account for situations like this.
This `fuzzy'  matching will be explored in future analysis of the dataset. 
Some prior work \cite{buzzell_modern_2024, buzzell_quantum_2024} has made use of Large Language Models in addition to manual classification in an effort to address the limitations of literal string searches as a classification tool. 
We explored use of a similar method in the early stages of this project but found that had its own shortcomings (e.g., a stochastic element to the analysis a reliance on arbitrary tuneable parameters to draw sharp distinctions) which made it unsuitable for this work.
One distinction between this and prior work is our focus on categorization rather than thematic analysis (where LLMs have proven more useful \cite{Comp_grounded_theory,caramaschi_uncovering}).

\section{Results}
We split the results into three distinct sections, each addressing one of the research questions. 
Section~\ref{sec:programs} will address results related to QISE programs. 
We then go onto discuss the entire dataset of courses from all disciplines in section~\ref{sec:all_courses}. 
The final set of results will focus on the courses that are offered in physics in section~\ref{sec:phy_courses}. 
We have also made all of these data on courses and programs, as well as a subset of the analysis presented herein, publicly available \cite{streamlit}.

\subsection{QISE programs}
\label{sec:programs}
In this section, we present results relevant to \textbf{RQ1}: What are the characteristics of QISE programs in higher education across the United States?

\subsubsection{Institutions with QISE programs}
\begin{figure*}[htb]
    \centering
    \includegraphics[width=\linewidth]{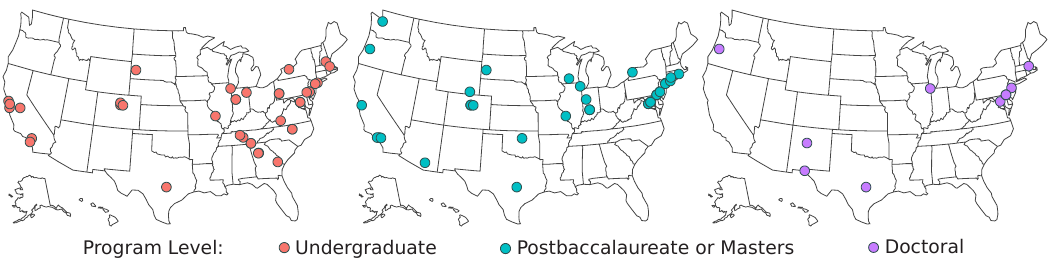}
    \caption{Maps of post-secondary institutions in the US with QISE programs. Each map includes all program types identified in this study (full degrees, certificates, minors, or tracks/concentrations/emphases) at their respective levels (undergraduate, post-baccalaureate or master's, and doctoral).}
    \label{fig:prog_map}
\end{figure*}

Across the sample of 1456 institutions, we have identified 61 US institutions with 89 distinct QISE programs, as some institutions have multiple programs (e.g., an undergraduate minor and a master's degree). 
These institutions, shown in Fig.~\ref{fig:prog_map}, are split by level of program (BS, MS, PhD). 
There are 28 states that do not have a single institution with a QISE program at any level. 
The largest concentration of programs are located in the Northeast region, Midwest (centered around Chicago), and Southern California. 
Aside from California, there is a notable lack of undergraduate programs in the western US.
PhD programs, in comparison to both undergraduate and masters level programs, are fewest in number, and narrowest in geographic distribution. 
Consistent with results from \citet{Meyer2024}, the more rural parts of the US are notably lacking access to QISE programs. 
Among institutions with QISE  programs, 54 are PhD granting, 5 are master's granting, 2 are four-year, and none are two-year institutions. 
PhD-granting institutions make up only 25\% of our sample of 1,458 institutions, yet they account for 89\% of those offering QISE programs.
This concentration aligns with prior landscape studies \cite{Cervantes2021, Meyer2024}.

Only 9 of the institutions with QISE programs are MSIs, despite there being 711 MSIs nationwide, all included in our sample. 
Notably, 8 of these are classified as research-intensive, reinforcing the idea that institutional resources strongly influence QISE program availability, regardless of MSI status.
Additionally, all MSIs with QISE programs are enrollment-based HSIs.

\subsubsection{Program levels and disciplines}

\begin{figure}[htb]
    \centering
    \includegraphics[width=\linewidth]{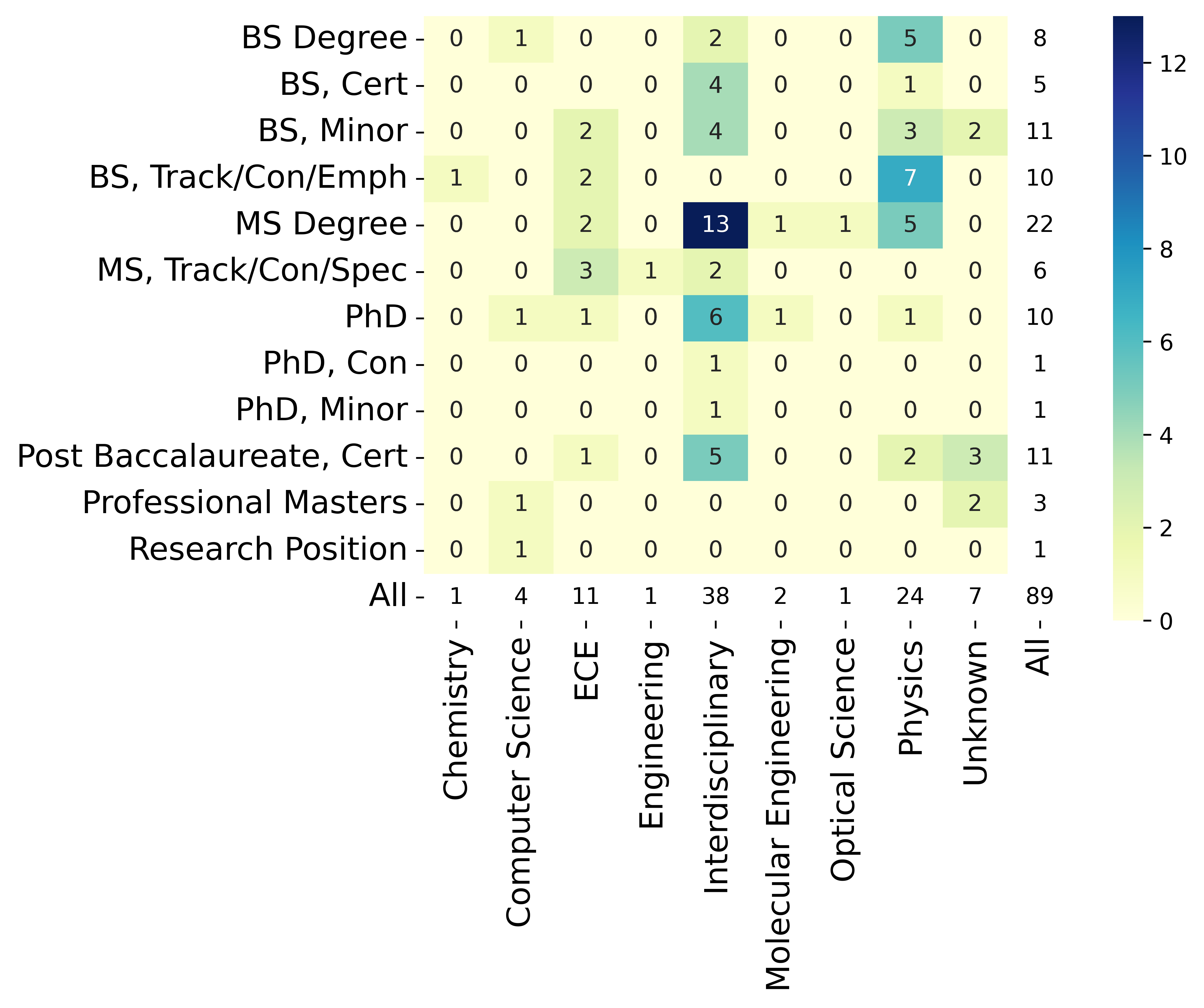}
    \caption{Heatmap showing the details of different QISE programs in the US. The vertical axis has information on the level and type of program being offered. The horizontal axis labels the discipline in which the program is offered.}
    \label{fig:program_crosstab}
\end{figure}

We examined QISE programs at the intersection of academic level and discipline, as shown in the heatmap in Fig.~\ref{fig:program_crosstab}. 
QISE programs are predominantly interdisciplinary (38), with significant contributions from physics (24) and electrical and computer engineering (11). 
Most QISE programs are offered at the graduate level, with no associate's degrees, 8 bachelor's programs, 24 master's programs, and 10 PhD programs.

We identified 11 bachelor's-level and 1 PhD-level QISE minor.
Minors are the most accessible QISE program type, as they allow students from various majors to enroll. 
In contrast, tracks, concentrations, foci, and emphases are structured within specific majors and require a defined set of QISE-related courses. 
We identified 10 bachelor's-level, 6 master's-level, and 1 PhD-level program of this type. 
Unlike minors, these program structures are limited in reach since they are restricted to students within a particular major.

Across both undergraduate and graduate levels, there are 16 different certification programs focused in QISE. 
They are largely focused on coursework and aim to provide participants with a base level of QISE knowledge upon which deeper understanding or practical skills could be built. 

\subsubsection{Requirements for QISE programs}

For some QISE-specific programs, admission requirements closely resemble those of traditional post-secondary education programs.
For bachelor's programs in QISE, students must first gain admission to the institution and, in some cases, to the specific major.
However, there are generally no additional requirements beyond those necessary for other majors at the institution.
For both undergraduate and graduate tracks or concentrations, students must typically be majors in the department offering the program. 
Minors, on the other hand, often accept students from a variety of major programs.
Master's and PhD programs in QISE are generally limited to applicants with an undergraduate STEM degree. 
Their admission application materials align with those of other graduate programs, typically including a CV, letters of recommendation, and standardized test scores, such as the GRE or English proficiency exams.
 
\subsection{Quantum and QISE in all recorded courses}
\label{sec:all_courses}
This section presents results relevant to \textbf{RQ2}: What is the distribution of courses that include quantum and/or QISE across different disciplines, modalities, and institution types? We recorded 8456 total courses whose title or description includes the word `quantum.' 

\begin{figure*}[htb]
    \centering
    \includegraphics[width=.9\linewidth]{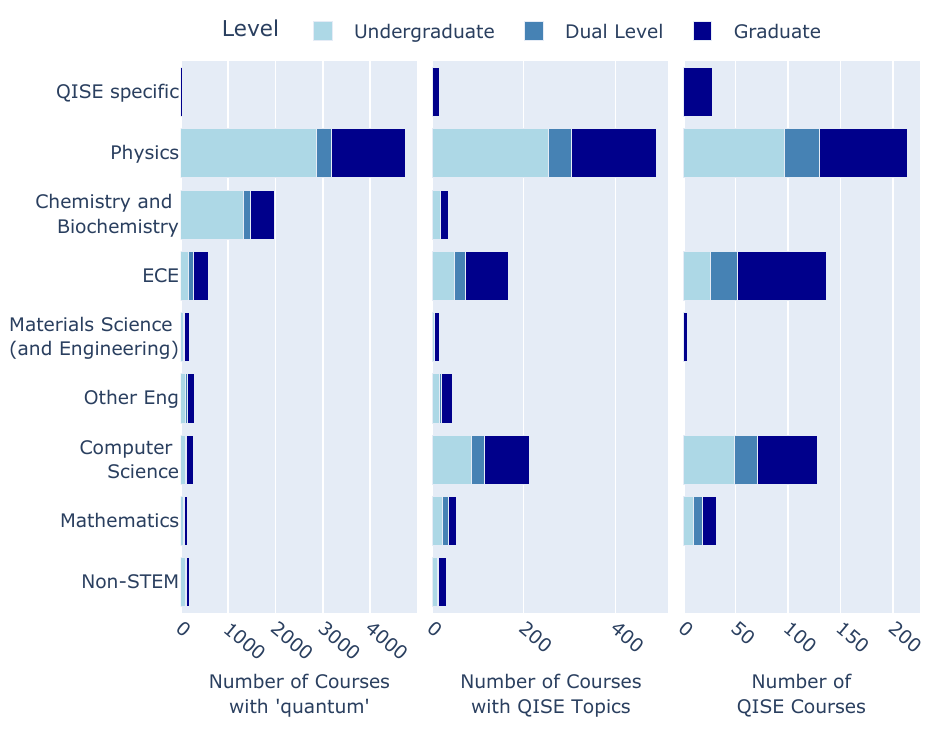}
    \caption{Three plots showing the number of courses, from different subsets of the data, offered within different disciplines. `All courses with quantum,' refers to the entire data set. `Courses with QISE topics' refers to the number of courses whose descriptions contained a QISE keyword (recall that `quantum' is not a QISE  keyword; QISE is a narrower category than `quantum'). QISE courses were categorized based on their titles and descriptions and were determined to be focused on primarily QISE content. Numerical values for these bars can be found in Appendix~\ref{chart_values}.}
    \label{fig:courses_by_discipline}
\end{figure*}

\subsubsection{What disciplines are covering `quantum' in their courses?}
A summary of the disciplines with courses whose titles or descriptions contain `quantum' is shown in Fig.~\ref{fig:courses_by_discipline} (Left). 
This plot reports the number of courses that were listed in each discipline. 

Cross-listed courses are counted multiple times in this representation, as it is often unclear which department should be considered primary. 
As a result, the total count in the leftmost plot sums to 8,652 rather than the 8,456 unique courses recorded.

We present a more detailed analysis of the approximately 4,700 physics courses---comprising nearly 60\% of all recorded courses---in Section~\ref{sec:phy_courses}. 
Notably, 1,994 courses (24\%) originate from chemistry and biochemistry departments, making this the second-largest discipline with courses mentioning quantum concepts.

Overall, most recorded courses are at the undergraduate level. However, outside of physics and chemistry, most disciplines offer more graduate-level than undergraduate courses on quantum topics. This suggests that, while quantum-related instruction is concentrated in undergraduate physics and chemistry courses, other fields tend to introduce quantum concepts primarily at the graduate level.

\subsubsection{What subset of courses cover QISE topics? }
Figure~\ref{fig:courses_by_discipline} (Center) shows the number of courses in different disciplines that had QISE topics specifically mentioned in their course descriptions. 
Many of these courses are not specifically about QISE and may vary considerably in the extent to which they cover QISE; these courses do have at least one of the QISE keywords (e.g., quantum computing) listed in Appendix~\ref{keywords}.
Although physics remains the single largest group of courses with QISE content, electrical and computer engineering and computer science become more prominent, and chemistry plays a comparatively smaller role..
Computer science and electrical and computer engineering courses with `quantum' cover QISE topics at higher rates than most other disciplines as well.
Approximately 82\% of computer science courses and 30\% electrical and computer engineering courses recorded cover QISE topics. 
Only about 2\% of chemistry courses that mention `quantum' contain QISE topics in their descriptions. 
This is the smallest proportion of courses covering QISE topics  to courses mentioning `quantum' from any discipline. 

\subsubsection{What subset of recorded courses are specific to QISE? }
\label{sec:all_qise_courses}
Across all disciplines, we have identified 529 distinct QISE courses 
(about 6\% of all recorded `quantum' courses). 
The distribution of these courses among disciplines is shown in Fig.~\ref{fig:courses_by_discipline}.
These courses are offered primarily in physics, computer science, and electrical and computer engineering. 

By comparing the total number of quantum courses in each discipline to the number of QISE-specific courses, we can assess the relative prevalence of QISE within different fields. 
In computer science, for example, 49\% of courses that mention `quantum' in their title or description are QISE-focused. 
This proportion is 23\% for both mathematics and electrical and computer engineering.
Although physics is the largest contributor to QISE course offerings, only 5\% of physics courses that cover `quantum' are specifically QISE-focused. 
Notably, chemistry is entirely absent from QISE course offerings; none of the nearly 2,000 recorded chemistry courses are classified as QISE-focused.

The most common QISE courses focus on quantum computing and quantum information, accounting for 418 out of 514 recorded courses.
Courses on the implementation of quantum technology form the next largest category, with 70 courses.
All other QISE course categories (see Fig. \ref{tab:course_categorization}) have fewer than 15 individual courses, highlighting the strong emphasis on computing and information within QISE education. 

\subsubsection{What courses that mention quantum contain a lab component?}
In our data set, we have identified 359 courses that contain a lab component or are full lab courses.  
To identify these courses, we searched course titles and descriptions for keywords such as "lab-" (a stem for laboratory) and "exp-" (a stem for experiment/al) to flag potential lab courses. 
These courses were then manually analyzed to confirm their lab status and to identify  for QISE topics (i.e., those shown in Fig. \ref{fig:QISE_topic_prevalence}), and any course containing at least one such topic was tagged as QISE.

Lab courses without QISE topics primarily covered concepts typically associated with theory-based instruction, such as measurement, wave-particle duality, and tunneling. These were predominantly introductory physics and chemistry courses, as well as physical chemistry and modern physics courses. The identified lab courses spanned a wide range, from introductory undergraduate courses to graduate-level offerings.

There were only 18 labs or courses with labs that contained experiments relevant to QISE. 
Of these, 13 were offered in physics departments, two of which were co-listed with electrical and computer engineering, and one of which was co-listed with an optics department. 
Other QISE lab courses were offered in electrical and computer engineering (2, 1 co-listed computer science), Optics (1), and dedicated QISE programs (2).

Five of these courses were explicitly focused on exclusively quantum computing. 
Titles ranged from ``Introduction to Quantum Computation'', to ``Experimental Methods in Quantum Computing.''
The foci of courses with experimental QISE components varied somewhat. 
Some courses focused on experiments that utilized or demonstrated the fundamental quantum physics used in quantum computation such as entanglement, superposition, and measurement. 
Others  emphasized primarily giving students hands-on experience with various quantum computing hardware platforms such as single photons, electron spins (Nitrogen vacancy centers in diamond), and superconducting qubits. 
One quantum computing lab description emphasized providing students with experience working with some technologies that can enable quantum computing such as high-performance analog electronics and cryogenic and vacuum techniques. 
Cryogenic and vacuum techniques were also prominently featured in the description of a more general course titled ``Quantum Technology Systems.''

Optical techniques for quantum technology were present in the descriptions of seven of the QISE courses with lab components, two of which were also among the quantum computing focused labs. 
Two other lab courses had somewhat general titles (``Quantum Lab'' and ``Experimental Quantum Information''). 
Another of these courses, ``Quantum Photonics and Communications,'' discussed having students assemble and test a commercially available quantum encrypted networks based on entangled photon pairs.
There were additionally two courses on ``Quantum and Nano Optics'' that in addition to discussing the use of optical methods for realizing quantum technologies, also discussed applications such as biotech and medicine.
Other common topics among the courses with optical techniques in their descriptions included lasers, ion trapping, entanglement, and single photon measurements.
 
Finally, one two-semester course sequence, while not a traditional lab course, is designed to give students first-hand experience with the quantum industry (see \cite{Oliver_qforge_2025}). 
This is achieved through partnerships with local industry collaborators on a industry-defined experimental project.

\subsubsection{How available are quantum courses?}
Across the whole sample, we found 484 (33\%) institutions that do not have a single course whose title or description mentions `quantum.' 
A cross-tabulation of those institutions' MSI statuses and Carnegie classifications is shown in Table~\ref{Tab:MSI_Carn_no_courses}.
Approximately 40\% of MSIs lack any courses with a title or description mentioning quantum, compared to 24\% of non-MSIs in this sample. 
Additionally, 56\% of two-year institutions and 44\% of four-year institutions offer no courses that reference quantum topics.
However, it is important to note that many two-year institutions were included in the sample due to our MSI criterion, which may have resulted in over-representation of two-year institutions.
Approximately 18\% of master's-granting institutions and 4\% of PhD-granting institutions do not have any courses mentioning quantum.
Across all four of these Carnegie classification groupings, there is a statistically significant difference (p<.00001) between the number of institutions with and without courses mentioning `quantum' and the corresponding Cramers V (V=0.413) indicates a medium effect size.
While it may seem surprising that some PhD-granting institutions lack courses with quantum in their title or description, several factors could explain this. 
Some universities offer primarily STEM bachelor's degrees, but have graduate programs only in a limited number of non-STEM fields. 
Similarly, other institutions have a small selection of STEM majors or a limited number of academic programs overall.

\begin{table}[htb]
\caption{Number of institutions with no courses whose title or description contained `quantum.' The counts are broken down by MSI status and Carnegie classification}
\begin{tabular}{c|c|c|cc}

\textbf{Carnegie} & \textbf{MSI-No} & \textbf{MSI-Yes} & \textbf{Sum} & \makecell[c]{\% of Institutions \\ in Sample} \\
\hline
\textbf{Two-Year} & 111 & 173 & 284 & 56\%\\
\textbf{Four-Year} & 37 & 82 & 119 & 44\% \\

\textbf{Master's} & 35 & 27 & 62 & 19\%\\

\textbf{PhD} & 12 & 7 & 19  & 4\%\\
\hline
\textbf{Sum} & 195 & 289 & 484 & - \\

\makecell[c]{\% of Institutions \\ in Sample} & 40\% & 24\%  & - & 33\%\\

\end{tabular}
\label{Tab:MSI_Carn_no_courses}

\end{table}

A total of 179 institutions (12\% of our sample) offer only one course with quantum in its title or description. 
These courses are distributed as follows: 40 (22\%) in general physics, 31 (17\%) in modern physics, 25 (14\%) in general chemistry, 22 (12\%) in physical chemistry, 20 (11\%) in quantum mechanics, and a single course (0.6\%) in introductory quantum computing. 
The remaining ~25\% include courses in nuclear physics, atomic physics, special topics, or imaging.

On average, non-MSIs offer 7.4 courses mentioning quantum, while MSIs offer an average of 3.9 per institution. 
However, this difference is less pronounced when comparing institutions within the same Carnegie classification (see Table ~\ref{tab:avg_qm_per_carn}).

\begin{table}[]
 \caption{Mean number of courses containing `quantum' at an institution of a given Carnegie classification and MSI status.}
    \centering
    \begin{tabular}{c|c|c}
        \textbf{Carnegie} & \textbf{MSI} & \textbf{Non-MSI} \\
        \hline
        \textbf{PhD-Granting} & 16 & 16 \\
        \textbf{Masters} & 4 & 4 \\
        \textbf{Four-Year} & 2 & 3 \\
        \textbf{Two-Year} & 1 & 1 \\
    \end{tabular}

    \label{tab:avg_qm_per_carn}
\end{table}

Of the 13 QISE-focused courses with lab components offered in physics departments, 11 of them are at PhD-granting institutions, 1 is at a master's granting institution, and another is at a four-year institution. 
This highlights a potential barrier to the implementation of QISE lab work. 
The cost to implement a QISE lab may be placing them out of reach of lower-resourced institutions. 
We suspect that these offerings are dictated in large part by the resources (funding, space, time)  and available faculty expertise of the institutions in which they are housed.

\subsection{Physics courses}
\label{sec:phy_courses}

In this section, we will address \textbf{RQ3}: Within physics departments, in what ways does the distribution by level, modality, type of course vary for quantum and QISE topics?
To address \textbf{RQ3}, we examined the subset of recorded courses that were listed in physics departments. 
These courses are grouped into categories (as discussed in Sec. \ref{sec:methods}) and examined for which QISE topics are being taught. 
We also report on the prerequisites for QISE courses listed in physics departments. 

\subsubsection{What types of physics courses include `quantum'?}
We identified 4740 quantum courses listed in physics departments, of which $\sim 4.5$\% (214) are QISE courses. 

\begin{figure*}[htb]
    \centering
    \includegraphics[width=\linewidth]{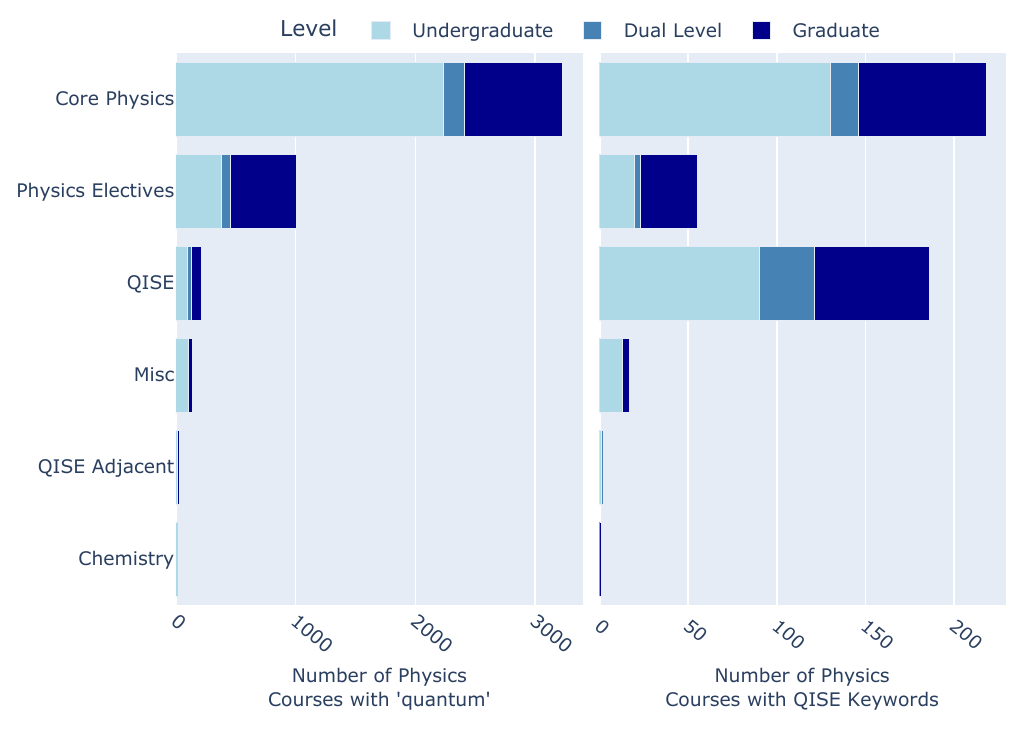}
    \caption{Number of physics courses of different types at different levels.}
    \label{fig:quantum_physics_course_types}
\end{figure*}

A summary of the number of each course type at different levels can be seen in Fig.~\ref{fig:quantum_physics_course_types}.
Within the set of physics courses mentioning `quantum,' 3229 ($\sim$70\%) fit into the category we refer to as the core physics curriculum. 
Core physics courses are defined as courses that \textit{most} physics majors would take at either the graduate or undergraduate level 
(introductory physics, modern physics, classical mechanics,  electricity and magnetism, statistical mechanics/thermodynamics, and quantum mechanics).
Approximately 50\% of the core physics courses that mention quantum are quantum mechanics and $\sim$20\% are modern physics.
Another $~$20\% of courses were categorized as physics electives, which may not be offered at every institution and are likely not taken by all physics majors at those institutions where they are offered. These courses include topics such as group and field theories, physics of materials, nuclear and particle physics, optics, computational physics, space science, nano-, physics of devices, atomic and molecular, biophysics, as well as  more specialized topics (chaos theory, information theory, and QED).
We also categorized miscellaneous physics courses, which include courses intended for non-majors, seminars, and special topics courses. These courses could not be assigned to more specific categories due to their broad scope or lack of detailed descriptions.

Finally, similar to the trend seen in the set of all QISE courses (see section \ref{sec:all_qise_courses}), the majority of QISE courses listed in physics are focused on quantum computing and information (179/214).
Courses on the implementations of quantum technology represent the next largest group (23/214) and there are fewer than 5 courses from each of the remaining types (see Fig. \ref{tab:course_categorization}).

\subsubsection{What types of physics courses include QISE topics?}
\label{sec:types of physics with qise topics}
To identify QISE topics in course descriptions, we began by using the collection of the most common topics found in introductory QIS courses, as identified by \citet{meyer_intro}.  Any topic reported in at least 50\% of the courses in the \citet{meyer_intro} study was included in the list of topics to search for in course descriptions. This search was conducted by looking for exact matches of specific words or phrases in the course descriptions.

A summary of the number of courses of each type separated by level is presented in Fig.~\ref{fig:quantum_physics_course_types}.
In addition to the 186 QISE-specific courses that resulted from this search, 254 core physics courses also mentioned QISE topics in their descriptions. 
Of these core physics courses, 170 (61\%)  were standard quantum mechanics courses; these 170 courses are 11\% of the 1545 standard quantum mechanics courses we recorded. 
The remaining 39\% of core physics courses consisted of modern physics (13\%), optics (8\%), statistical mechanics (3\%), and 14 other course types (<2\%) each.

We also counted the number of courses mentioning specific QISE keywords from Meyer et al., ( See Sec.~\ref{sec:types of physics with qise topics}), as shown in Fig.~\ref{fig:QISE_topic_prevalence}. 
Most topics include multiple related keywords, which is fully described in  Appendix~\ref{keywords}.
This breakdown highlights several key points.
It underscores the relative exclusivity of certain topics to QISE courses. 
For example, quantum algorithms and cryptography are found almost exclusively in QISE courses. 
This is particularly notable given that discussions of quantum gates and advanced theory are also present in many non-QISE course descriptions. 
Finally, these findings provide evidence of the integration of QISE topics into non-QISE courses.

\begin{figure}[htb]
    \centering
    \includegraphics[width=\linewidth]{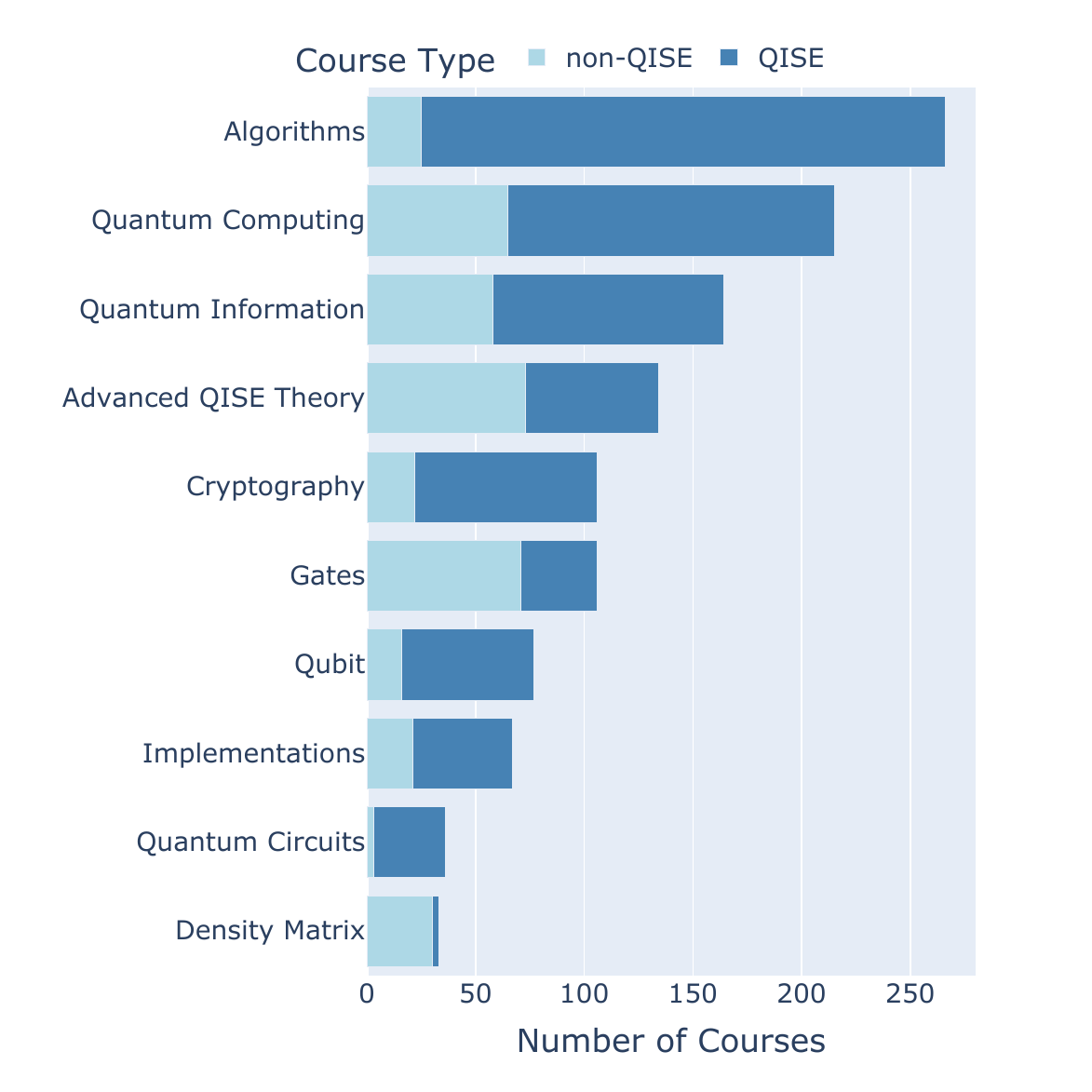}
    \caption{Number of QISE and non-QISE courses offered in physics that include some QISE  topic(s) in their descriptions.}
    \label{fig:QISE_topic_prevalence}
\end{figure}

\subsubsection{What are the prerequisites for QISE courses in Physics? }
\begin{figure}[htb]
    \centering
    \includegraphics[width = \linewidth]{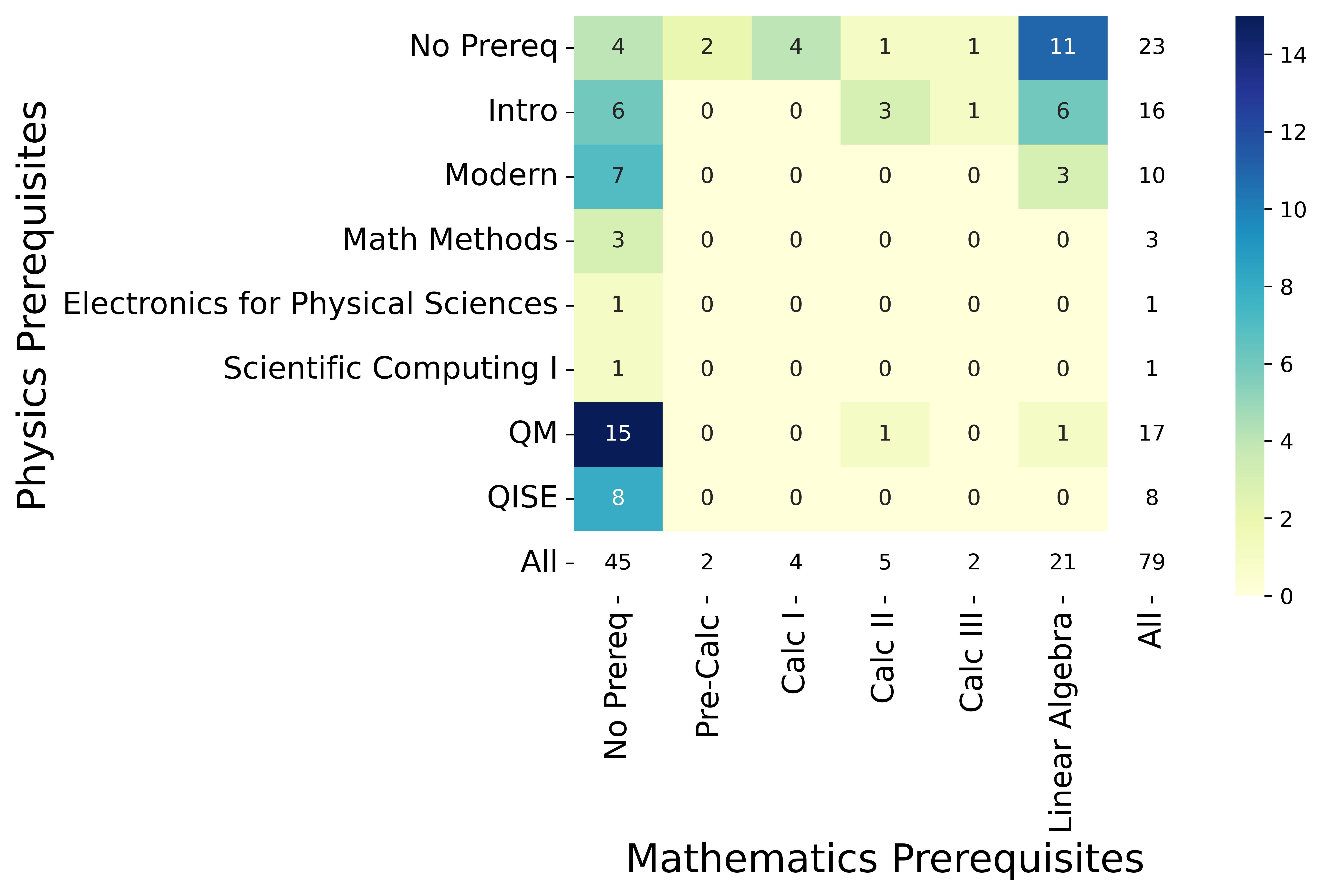}
    \caption{Heatmap showing the number of undergraduate QISE courses offered in physics with different combinations of physics and mathematics prerequisites. Both axes are organized by level of sophistication of the prerequisites with the lowest levels at the top row and leftmost column. }
    \label{fig:prereq_heatmap}
\end{figure}

We were able to access prerequisite information for 133 (61\% of 214 total) QISE courses listed in physics at both the undergraduate and graduate level.
Mathematics and physics are the two most common prerequisite subjects. 
Figure~\ref{fig:prereq_heatmap} depicts the unique combinations of different physics and mathematics prerequisites for the 79 undergraduate QISE courses listed in physics. 
We did not include graduate courses in this representation given the assumed prerequisite of a four-year degree.
In the case of multiple mathematics or physics prerequisites, this representation considers only the highest level course requirement from each discipline. 
Beyond the mathematics and physics prerequisites that are shown in Fig.~\ref{fig:prereq_heatmap}, 8 courses have electrical and computer engineering prerequisites, 13 have computer science prerequisites, 2 have mechanical engineering prerequisites, and 1 has a chemistry prerequisite.

There are a few notable features of Fig.~\ref{fig:prereq_heatmap}. 
Courses with no physics prerequisite (top row) and/or no math prerequisites (left column) are particularly interesting. 
There are 4 undergraduate courses with no mathematics or physics prerequisites, and an additional 19 courses with some mathematics prerequisite and no physics prerequisite. 
These courses could be indicative of efforts to introduce students to QISE topics earlier in their undergraduate education (e.g., \cite{economou_hello_2022}).
This interpretation is supported by the data on course numbers, which showed that among these 79 undergraduate courses, 26 had course numbers that would imply they are introductory (e.g., course numbers beginning in 1- or 2- hundred/thousand depending on the institution).

Similarly, several courses do not require mathematics above calculus III (13 courses) or physics beyond introductory physics (16 courses).
We must note, however, that although a course may not list a mathematics or physics prerequisite explicitly, some prerequisites may be `hidden.' 
For example, a QISE course may require quantum mechanics, but not have a specific mathematics prerequisite.
In such cases, the quantum mechanics course (or its associated prerequisites) could have involved more advanced mathematics, such as differential equations or linear algebra.
The most common physics and mathematics prerequisites are quantum mechanics and linear algebra respectively.
Approximately 22\% (17/73) of QISE courses require quantum mechanics and 27\% (21/79) of QISE courses require linear algebra.

\subsubsection{What is the availability of physics courses that include quantum and QISE topics?}

One way we can measure the availability of courses is by averaging the number of coursers across institutions. 
In this case, we averaged the number of physics courses including quantum at institutions of a given Carnegie classification. 
A summary of this analysis can be found in Table~\ref{tab:num_phy_courses}.
This metric gives some indication of how accessible these topics are to students at different institutions types. 

Roughly half of two-year institutions offer a physics course that mentions `quantum.' 
However, only about 1 in 100 two-year institutions have physics courses that address QISE topics. 
Four-year institutions, on average, offer between 1 and 2 courses that mention `quantum,' with only about 1 in 10 offering a course that covers QISE topics. 
Master's-granting institutions typically offer between 2 and 3 courses that mention `quantum,' and, similarly, about 1 in 10 include a course on QISE topics. 
PhD-granting institutions, on average, have more than 8 physics courses that mention `quantum' and at least one course that covers QISE topics. 

When we examine physics courses with `quantum' courses by level, we see that there are overall more undergraduate than graduate courses. 
This trend is particularly striking in core physics courses, where over two-thirds of the $\sim$3200 are at the undergraduate level (see Fig.~\ref{fig:quantum_physics_course_types}). 
This trend is similar for physics courses with QISE keywords, where approximately two-thirds of them are at the undergraduate level. 

\begin{table}[htb]
    \centering
        \caption{Average number of courses offered per institution in physics departments. QM refers to any course with a title or description that includes `quantum.' QISE refers to any course with a course description that includes one or more of the QISE topics in Fig.~\ref{fig:QISE_topic_prevalence}.}
    \begin{tabular}{c|c|c}
       \textbf{ Institution Type }  & \textbf{ QM } & \textbf{ QISE }  \\
       \hline
       \textbf{Two-Year}  & 0.51 & 0.01 \\
       \textbf{Four-Year} & 1.43 & 0.12 \\
       \textbf{Master's} & 2.42 & 0.14  \\
       \textbf{PhD} & 8.63 &  1.13 \\
    \end{tabular}
    \label{tab:num_phy_courses}
\end{table}

\section{Summary and Discussion}
This work aims to address three broad research questions. 
Here, we synthesize our results and discuss how collectively the outcomes of the analysis help to answer these questions.

\subsection{RQ1: What are the characteristics of QISE programs in higher education across the United States?  }
We examined 1456 degree granting institutions in the US, which encompassed all research intensive institutions, all MSIs, all institutions with relevant ABET accreditation, and the institutions that grant the most STEM degrees relevant to QISE within each state. 
We identified 61 institutions with QISE programs; this is an increase from the 34 institutions identified by \citet{Meyer2024}. 
All institutions identified as having QISE programs in \citet{Meyer2024} were also identified in this study, with the exception of one which has stopped being offered between their data collection (2019/2020) and our data (summer 2024).  
It is difficult to determine the extent to which this represents actual growth in the number of QISE programs being offered over time, or if it a result of searching a wider range of institutions. 
We suspect that it is a combination of both of these factors.

Consistent with previous findings, we found that QISE programs are highly concentrated in PhD granting institutions  (54/61 institutions). 
Only about 3\% of MSIs have QISE programs compared to 6\% of non-MSIs. 
Meyer et al.'s regression analysis suggests that the presence of QISE programs is largely explained by having access to more institutional resources \cite{Meyer2024}. 
They also note, and we agree, that while this may be the case, it does not mitigate the impact on students' access to QISE programs. 

Of the 89 individual programs offered between these institutions, the largest group of them is interdisciplinary, typically involving combinations of physics, electrical and computer engineering, and computer science. 
This suggests that the field of QISE may be shifting from purely physics to a broader range of disciplines. 
This is further evidenced by electrical and computer engineering, which, at both the undergraduate and master's level, is the second largest single field contributing to programs. 
Across the various program types, there are 34 QISE programs offered at the undergraduate level, which compares to $\sim$39 programs offered between master's and PhD levels. 
This may be indicative of institutions working to make QISE more accessible to students and address calls to support students' transition to the QISE workforce.

\subsection{RQ2: What is the distribution of courses that include quantum and/or QISE across different disciplines, modalities, and institution types? }
Across all institutions, we identified 8456 courses with `quantum' in their title or description. 
Although more than half of these came from physics, there were also nearly 2000 courses from chemistry departments. 
When we focused on QISE focused courses, we saw a different distribution. 
The majority of QISE courses were taught in physics, computer science, and electrical and computer engineering. 
Despite the potentially significant implications of quantum computing and other quantum technologies on fields like chemistry, we did not identify any QISE-specific courses being offered in chemistry. 
For a considerable number of institutions, the only access students have to quantum is through their chemistry coursework. 
This implies that chemistry could be leveraged as a way to introduce a greater variety of students to QISE topics.

We also examined the modality of QISE courses and found that very few include QISE lab components.
While industry needs include many experimental skills and knowledge \cite{hughes2022}, the fact that the $\sim$98\% of QISE courses have only theoretical components suggests a potential gap between educational offerings and job-relevant knowledge and skills. 
Addressing this gap is a complex challenge. 
Hands-on QISE labs can be resource intensive and there is currently a lack of adaptable course materials for new courses to build from. 
These scarce QISE lab courses are also very disproportionately represented at non-MSI schools and schools with graduate programs. 
Again, since experimental skills are regarded as an important qualification for the industry workforce \cite{hughes2022}, this unequal distribution suggests that our two-year schools and our MSIs may be leaving certain populations at an increased disadvantage when it comes to participating in the QISE workforce. 
There is a potential need for the development of modular, affordable QISE lab activities that could be deployed at lower-resourced and teaching-focused institutions.

The emerging nature of the field, coupled with the absence of shareable course materials, often makes hands-on QISE education resemble academic research more than laboratory instruction and is thus not scalable to engage a large number of students. 
This is further evidenced by the diversity of topics and experiments in the limited number of QISE lab courses we identified, underscoring the need for greater accessibility in experimental quantum education through scalable lab courses.

The availability of courses teaching quantum and QISE varies noticeably with certain institutional characteristics. 
Consistent with the results of Meyer et al. \cite{Meyer2024}, our data show that QISE coursework is highly concentrated in PhD granting institutions. 
One third of institutions, composed of primarily two- and four-year institutions, do not have a single course that mentions `quantum.' 
Although initiatives such as the NQI Act \cite{NQI}, the QIST Workforce Development NSP \cite{NSP}, and the CHIPS and Science Act \cite{CHIPS} aim to strengthen the quantum workforce, these data suggest that a crucial intermediary step is expanding access to quantum-related coursework. 

\subsection{RQ3:  Within physics departments, in what ways does the distribution by level, modality, and type of course vary for quantum and QISE topics? }

Physics departments offer more courses covering quantum-related content than all other disciplines in this study combined, which is unsurprising considering quantum mechanics plays a central role in contemporary physics curriculum, as demonstrated by its presence in a variety of different course types across different levels. 
Within physics, there are more non-QISE courses covering topics relevant to QISE than there are dedicated QISE courses. 
Most of these non-QISE courses are a part of the core curriculum for physics majors (bachelor's degree), such as modern physics (often taken in year 2) and quantum mechanics (often taken in year 3 or 4). 
Even if an institution cannot dedicate the resources to develop or implement an entire course in QISE, there is the possibility of implementing QISE content into existing courses. 
Some of the most common QISE topics/keywords included in non-QISE course descriptions are related to quantum gates, entanglement, quantum information, and quantum computing. 
This is consistent with Buzzell et al.'s findings that quantum computing was a common topic referenced in syllabi for quantum mechanics courses \cite{buzzell_modern_2024}. 

The alignment between core physics coursework and QISE topics raises questions about how prerequisite structures shape students' access to dedicated QISE courses. Prerequisite data we have on QISE courses within physics can be broken into three primary categories.
The first is comprised of QISE course that do not have any physics prerequisites (23 courses). 
A comparable number only require introductory or modern physics (26 courses).
On the other end of the spectrum, there are 23 courses that require either quantum mechanics or another QISE course. 
While quantum mechanics still serves as a gateway to QISE in some cases, the distribution of prerequisites suggests a growing effort to make QISE instruction more accessible to a broader range of students.

\section{Conclusion}
This work contributes to a broader effort to develop a more comprehensive picture of the current landscape of quantum and QISE courses and programs in higher education.
We hope that by making these data publicly available there will be opportunities for instructors, program and course developers, employers, and policy makers to note where QISE education is present and where more directed investment and development is necessary. 

While the number of QISE programs in the US is increasing, they remain concentrated in research-intensive institutions and specific geographic regions.
QISE programs are inherently interdisciplinary, drawing from physics, engineering, computer science and other fields. 
Notably, disciplines outside of physics have already begun creating dedicated  QISE courses or incorporating QISE topics in existing courses. 
Given the interdisciplinary nature of QISE, physics departments have an unique opportunity to offer introductory QISE courses as service courses for STEM majors across the university. 
In the absence of dedicated course offerings, computer science and electrical and computer engineering programs are particularly well-positioned to integrate QISE content into existing curricula, thereby serving as models for effective curricular incorporation across disciplines.

For institutions where developing new QISE courses is not feasible, embedding QISE content into existing courses, such as modern physics, offers a practical alternative. 
This strategy could significantly expand access to QISE education, particularly at two- and four-year institutions, helping to broaden participation in the field.

\section{Future Work}

Some limitations of this work stem from the constraints of publicly available data and the inferences that can be drawn from it. 
To gain a deeper understanding of the QISE education landscape, qualitative data, such as interviews with faculty and program developers, could provide valuable insights into the challenges and opportunities they encounter in developing and implementing these programs.

Even when QISE courses are available, it remains unclear how well current academic structures prepare students for careers in government and industry. 
Our future work will explore perspectives from both educators and industry members to identify alignment between workforce needs and educational goals. 
By fostering this synergy, we aim to support a more vibrant QISE educational ecosystem.

\acknowledgments{This material is based on work supported by the National Science Foundation under Grant Nos. PHY-2333073 and PHY-2333074. The authors thank the members of the CASTLE group for their feedback and insight on the project and the manuscript. We thank the members of the CU PER group who provided feedback on the website where the data for this paper are publicly available and downloadable: quantumlandscape.streamlit.app. }

\onecolumngrid
\appendix
\section{Carnegie Simplification}
\label{carnegie_simp_sec}
\begin{table*}[htb]
    \centering
      \caption{Broad categories created from grouping Carnegie classifications. Due to the large number of individual Carnegie classifications, we combined them to create four broader categories. The details of which Carnegie classifications are in which simplified category are here.}
    \begin{tabular}{|c|m{12cm}|}
    \hline
         PhD Granting Institutions &
         \begin{itemize}
         \vspace{6pt}
             \item Doctoral Universities: Very High Research Activity (R1)
             \item Doctoral Universities: High Research Activity (R2)
             \item Doctoral/Professional Universities
         \end{itemize}\\
         \hline
         Master's Colleges and Universities &
         \begin{itemize}
                  \vspace{6pt}
             \item Master's Colleges and Universities: Larger Programs
             \item Master's Colleges and Universities: Medium Programs
             \item Master's Colleges and Universities: Small Programs
         \end{itemize}\\
         \hline
         Four-Year Institutions &
         \begin{itemize}
         \vspace{6pt}
             \item Baccalaureate Colleges: Arts \& Sciences Focus
             \item Baccalaureate Colleges: Diverse Fields
             \item Special Focus Four-Year: Faith-Related Institutions
             \item Special-Focus Four-Year: Engineering and Other Technology-Related Schools
             \item Baccalaureate/Associate's Colleges: Mixed Baccalaureate/Associate's
         \end{itemize}\\
         \hline
         Two-Year Institutions &
         \begin{itemize}
         \vspace{6pt}
             \item Associate's Colleges: Mixed Transfer/Career \& Technical-High Nontraditional
             \item Associate's Colleges: High Career \& Technical-Mixed Traditional/Nontraditional
             \item Associate's Colleges: Mixed Transfer/Career \& Technical-Mixed Traditional/Nontraditional
             \item Associate's Colleges: Mixed Transfer/Career \& Technical-High Traditional
             \item Associate's Colleges: High Transfer-High Nontraditional
             \item Associate's Colleges:High Career \& Technical-High Nontraditional
             \item Associate's Colleges: High Transfer-Mixed Traditional/Nontraditional
             \item Associate's Colleges:High Transfer-High Traditional
             \item Special Focus Two-Year: Health Professions
             \item Baccalaureate/Associate's Colleges: Associate's Dominant
         \end{itemize}\\
         \hline
    \end{tabular}
  
    \label{tab:carnegie_simp}
\end{table*}

\pagebreak
\section{Full Institutional Characteristics of Sample}
\label{sample_characteristics}
\begin{table}[htb]
\begin{tabular}{|ccccc|c|}
\hline
\multicolumn{5}{|c|}{\textbf{\begin{tabular}[c]{@{}c@{}}Institutional \\ Characteristics\end{tabular}}}                                                                                                                                                         & \multirow{2}{*}{\textbf{\begin{tabular}[c]{@{}c@{}}Number of\\  Institutions\end{tabular}}} \\ \cline{1-5}
\multicolumn{1}{|c|}{\textbf{\begin{tabular}[c]{@{}c@{}}Simplified \\ Carnegie\\ \end{tabular}}}   & \multicolumn{1}{c|}{\textbf{MSI}}         & \multicolumn{1}{c|}{\textbf{\begin{tabular}[c]{@{}c@{}}Top 10 \\ Bachelor's\\ \end{tabular}}} & \multicolumn{1}{c|}{\textbf{\begin{tabular}[c]{@{}c@{}}Top 5 \\ Associate's\end{tabular}}} & \textbf{\begin{tabular}[c]{@{}c@{}}ABET Accredited\\ CS or Eng\end{tabular}} &                                                  \\ \hline
\multicolumn{1}{|c|}{\multirow{15}{*}{Four-Year}}    & \multicolumn{1}{c|}{\multirow{7}{*}{No}}  & \multicolumn{1}{c|}{\multirow{3}{*}{No}}     & \multicolumn{1}{c|}{No}                      & Yes                 & 58                                               \\ \cline{4-6} 
\multicolumn{1}{|c|}{}                               & \multicolumn{1}{c|}{}                     & \multicolumn{1}{c|}{}                        & \multicolumn{1}{c|}{\multirow{2}{*}{Yes}}    & No                  & 5                                                \\ \cline{5-6} 
\multicolumn{1}{|c|}{}                               & \multicolumn{1}{c|}{}                     & \multicolumn{1}{c|}{}                        & \multicolumn{1}{c|}{}                        & Yes                 & 2                                                \\ \cline{3-6} 
\multicolumn{1}{|c|}{}                               & \multicolumn{1}{c|}{}                     & \multicolumn{1}{c|}{\multirow{4}{*}{Yes}}    & \multicolumn{1}{c|}{\multirow{2}{*}{No}}     & No                  & 34                                               \\ \cline{5-6} 
\multicolumn{1}{|c|}{}                               & \multicolumn{1}{c|}{}                     & \multicolumn{1}{c|}{}                        & \multicolumn{1}{c|}{}                        & Yes                 & 18                                               \\ \cline{4-6} 
\multicolumn{1}{|c|}{}                               & \multicolumn{1}{c|}{}                     & \multicolumn{1}{c|}{}                        & \multicolumn{1}{c|}{\multirow{2}{*}{Yes}}    & No                  & 4                                                \\ \cline{5-6} 
\multicolumn{1}{|c|}{}                               & \multicolumn{1}{c|}{}                     & \multicolumn{1}{c|}{}                        & \multicolumn{1}{c|}{}                        & Yes                 & 1                                                \\ \cline{2-6} 
\multicolumn{1}{|c|}{}                               & \multicolumn{1}{c|}{\multirow{8}{*}{Yes}} & \multicolumn{1}{c|}{\multirow{4}{*}{No}}     & \multicolumn{1}{c|}{\multirow{2}{*}{No}}     & No                  & 123                                              \\ \cline{5-6} 
\multicolumn{1}{|c|}{}                               & \multicolumn{1}{c|}{}                     & \multicolumn{1}{c|}{}                        & \multicolumn{1}{c|}{}                        & Yes                 & 11                                               \\ \cline{4-6} 
\multicolumn{1}{|c|}{}                               & \multicolumn{1}{c|}{}                     & \multicolumn{1}{c|}{}                        & \multicolumn{1}{c|}{\multirow{2}{*}{Yes}}    & No                  & 1                                                \\ \cline{5-6} 
\multicolumn{1}{|c|}{}                               & \multicolumn{1}{c|}{}                     & \multicolumn{1}{c|}{}                        & \multicolumn{1}{c|}{}                        & Yes                 & 1                                                \\ \cline{3-6} 
\multicolumn{1}{|c|}{}                               & \multicolumn{1}{c|}{}                     & \multicolumn{1}{c|}{\multirow{4}{*}{Yes}}    & \multicolumn{1}{c|}{\multirow{2}{*}{No}}     & No                  & 6                                                \\ \cline{5-6} 
\multicolumn{1}{|c|}{}                               & \multicolumn{1}{c|}{}                     & \multicolumn{1}{c|}{}                        & \multicolumn{1}{c|}{}                        & Yes                 & 2                                                \\ \cline{4-6} 
\multicolumn{1}{|c|}{}                               & \multicolumn{1}{c|}{}                     & \multicolumn{1}{c|}{}                        & \multicolumn{1}{c|}{\multirow{2}{*}{Yes}}    & No                  & 2                                                \\ \cline{5-6} 
\multicolumn{1}{|c|}{}                               & \multicolumn{1}{c|}{}                     & \multicolumn{1}{c|}{}                        & \multicolumn{1}{c|}{}                        & Yes                 & 1                                                \\ \hline
\multicolumn{1}{|c|}{\multirow{11}{*}{Master's}}      & \multicolumn{1}{c|}{\multirow{6}{*}{No}}  & \multicolumn{1}{c|}{\multirow{2}{*}{No}}     & \multicolumn{1}{c|}{No}                      & Yes                 & 91                                               \\ \cline{4-6} 
\multicolumn{1}{|c|}{}                               & \multicolumn{1}{c|}{}                     & \multicolumn{1}{c|}{}                        & \multicolumn{1}{c|}{Yes}                     & No                  & 3                                                \\ \cline{3-6} 
\multicolumn{1}{|c|}{}                               & \multicolumn{1}{c|}{}                     & \multicolumn{1}{c|}{\multirow{4}{*}{Yes}}    & \multicolumn{1}{c|}{\multirow{2}{*}{No}}     & Yes                 & 61                                               \\ \cline{5-6} 
\multicolumn{1}{|c|}{}                               & \multicolumn{1}{c|}{}                     & \multicolumn{1}{c|}{}                        & \multicolumn{1}{c|}{}                        & No                  & 35                                               \\ \cline{4-6} 
\multicolumn{1}{|c|}{}                               & \multicolumn{1}{c|}{}                     & \multicolumn{1}{c|}{}                        & \multicolumn{1}{c|}{\multirow{2}{*}{Yes}}    & Yes                 & 10                                               \\ \cline{5-6} 
\multicolumn{1}{|c|}{}                               & \multicolumn{1}{c|}{}                     & \multicolumn{1}{c|}{}                        & \multicolumn{1}{c|}{}                        & No                  & 7                                                \\ \cline{2-6} 
\multicolumn{1}{|c|}{}                               & \multicolumn{1}{c|}{\multirow{5}{*}{Yes}} & \multicolumn{1}{c|}{\multirow{2}{*}{No}}     & \multicolumn{1}{c|}{\multirow{2}{*}{No}}     & No                  & 76                                               \\ \cline{5-6} 
\multicolumn{1}{|c|}{}                               & \multicolumn{1}{c|}{}                     & \multicolumn{1}{c|}{}                        & \multicolumn{1}{c|}{}                        & Yes                 & 34                                               \\ \cline{3-6} 
\multicolumn{1}{|c|}{}                               & \multicolumn{1}{c|}{}                     & \multicolumn{1}{c|}{\multirow{3}{*}{Yes}}    & \multicolumn{1}{c|}{\multirow{2}{*}{No}}     & Yes                 & 14                                               \\ \cline{5-6} 
\multicolumn{1}{|c|}{}                               & \multicolumn{1}{c|}{}                     & \multicolumn{1}{c|}{}                        & \multicolumn{1}{c|}{}                        & No                  & 7                                                \\ \cline{4-6} 
\multicolumn{1}{|c|}{}                               & \multicolumn{1}{c|}{}                     & \multicolumn{1}{c|}{}                        & \multicolumn{1}{c|}{Yes}                     & Yes                 & 1                                                \\ \hline
\multicolumn{1}{|c|}{\multirow{11}{*}{PhD Granting}} & \multicolumn{1}{c|}{\multirow{6}{*}{No}}  & \multicolumn{1}{c|}{\multirow{2}{*}{No}}     & \multicolumn{1}{c|}{\multirow{2}{*}{No}}     & Yes                 & 61                                               \\ \cline{5-6} 
\multicolumn{1}{|c|}{}                               & \multicolumn{1}{c|}{}                     & \multicolumn{1}{c|}{}                        & \multicolumn{1}{c|}{}                        & No                  & 15                                               \\ \cline{3-6} 
\multicolumn{1}{|c|}{}                               & \multicolumn{1}{c|}{}                     & \multicolumn{1}{c|}{\multirow{4}{*}{Yes}}    & \multicolumn{1}{c|}{\multirow{2}{*}{No}}     & Yes                 & 170                                              \\ \cline{5-6} 
\multicolumn{1}{|c|}{}                               & \multicolumn{1}{c|}{}                     & \multicolumn{1}{c|}{}                        & \multicolumn{1}{c|}{}                        & No                  & 19                                               \\ \cline{4-6} 
\multicolumn{1}{|c|}{}                               & \multicolumn{1}{c|}{}                     & \multicolumn{1}{c|}{}                        & \multicolumn{1}{c|}{\multirow{2}{*}{Yes}}    & Yes                 & 5                                                \\ \cline{5-6} 
\multicolumn{1}{|c|}{}                               & \multicolumn{1}{c|}{}                     & \multicolumn{1}{c|}{}                        & \multicolumn{1}{c|}{}                        & No                  & 3                                                \\ \cline{2-6} 
\multicolumn{1}{|c|}{}                               & \multicolumn{1}{c|}{\multirow{5}{*}{Yes}} & \multicolumn{1}{c|}{\multirow{3}{*}{No}}     & \multicolumn{1}{c|}{\multirow{2}{*}{No}}     & Yes                 & 29                                               \\ \cline{5-6} 
\multicolumn{1}{|c|}{}                               & \multicolumn{1}{c|}{}                     & \multicolumn{1}{c|}{}                        & \multicolumn{1}{c|}{}                        & No                  & 23                                               \\ \cline{4-6} 
\multicolumn{1}{|c|}{}                               & \multicolumn{1}{c|}{}                     & \multicolumn{1}{c|}{}                        & \multicolumn{1}{c|}{Yes}                     & No                  & 1                                                \\ \cline{3-6} 
\multicolumn{1}{|c|}{}                               & \multicolumn{1}{c|}{}                     & \multicolumn{1}{c|}{\multirow{2}{*}{Yes}}    & \multicolumn{1}{c|}{\multirow{2}{*}{No}}     & Yes                 & 49                                               \\ \cline{5-6} 
\multicolumn{1}{|c|}{}                               & \multicolumn{1}{c|}{}                     & \multicolumn{1}{c|}{}                        & \multicolumn{1}{c|}{}                        & No                  & 4                                                \\ \hline
\multicolumn{1}{|c|}{\multirow{10}{*}{Two-Year}}     & \multicolumn{1}{c|}{\multirow{3}{*}{No}}  & \multicolumn{1}{c|}{No}                      & \multicolumn{1}{c|}{No}                      & Yes                 & 35                                               \\ \cline{3-6} 
\multicolumn{1}{|c|}{}                               & \multicolumn{1}{c|}{}                     & \multicolumn{1}{c|}{\multirow{2}{*}{No}}     & \multicolumn{1}{c|}{\multirow{2}{*}{Yes}}    & No                  & 130                                              \\ \cline{5-6} 
\multicolumn{1}{|c|}{}                               & \multicolumn{1}{c|}{}                     & \multicolumn{1}{c|}{}                        & \multicolumn{1}{c|}{}                        & Yes                 & 17                                               \\ \cline{2-6} 
\multicolumn{1}{|c|}{}                               & \multicolumn{1}{c|}{\multirow{7}{*}{Yes}} & \multicolumn{1}{c|}{\multirow{4}{*}{No}}     & \multicolumn{1}{c|}{\multirow{2}{*}{No}}     & No                  & 266                                              \\ \cline{5-6} 
\multicolumn{1}{|c|}{}                               & \multicolumn{1}{c|}{}                     & \multicolumn{1}{c|}{}                        & \multicolumn{1}{c|}{}                        & Yes                 & 12                                               \\ \cline{4-6} 
\multicolumn{1}{|c|}{}                               & \multicolumn{1}{c|}{}                     & \multicolumn{1}{c|}{}                        & \multicolumn{1}{c|}{\multirow{2}{*}{Yes}}    & No                  & 39                                               \\ \cline{5-6} 
\multicolumn{1}{|c|}{}                               & \multicolumn{1}{c|}{}                     & \multicolumn{1}{c|}{}                        & \multicolumn{1}{c|}{}                        & Yes                 & 7                                                \\ \cline{3-6} 
\multicolumn{1}{|c|}{}                               & \multicolumn{1}{c|}{}                     & \multicolumn{1}{c|}{\multirow{3}{*}{Yes}}    & \multicolumn{1}{c|}{No}                      & No                  & 2                                                \\ \cline{4-6} 
\multicolumn{1}{|c|}{}                               & \multicolumn{1}{c|}{}                     & \multicolumn{1}{c|}{}                        & \multicolumn{1}{c|}{\multirow{2}{*}{Yes}}    & No                  & 1                                                \\ \cline{5-6} 
\multicolumn{1}{|c|}{}                               & \multicolumn{1}{c|}{}                     & \multicolumn{1}{c|}{}                        & \multicolumn{1}{c|}{}                        & Yes                 & 1                                                \\ \hline
\end{tabular}
\end{table}
\pagebreak

\section{Discipline Binning}
\label{dept_simp}
\begin{table*}[htb]
    \centering
      \caption{Thematic groupings of Department/discipline categories. Department/discipline names are not consistent across different institutions, so binned similar departments according to the scheme in this table. The left column are the disciplines reported on in the body of the paper. The right column reports the individual departments/disciplines included in the larger discipline according to how they are reported in the course catalogs where we retrieved the course information.}
    \begin{tabular}{|K{2.5cm}|m{7.5cm}m{7.5cm}|}
    \hline

    QISE-Specific & 
    \begin{itemize}
        \vspace{6pt}
        \item Quantum Science and Technology
        \item Quantum materials science and engineering
        \item Quantum Computing
    \end{itemize}&
    \begin{itemize}
        \item Quantum Science and Engineering
        \item Quantum information sciences
    \end{itemize}
    \\
\hline 
    
    Physics &
    \begin{itemize}
        \vspace{6pt}
        \item Engineering Physics
        \item Biophysics
        \item Applied \& Engineering Physics
        \item Applied Physics
        \item Physical Science
        \item Applied Physics and Materials Science
        \item Physics and Energy Science
    \end{itemize}&
    \begin{itemize}
        \item Physics and Engineering Physics
        \item Applied Physics and Mathematics
        \item Physics for Educators
        \item Health Physics
        \item Physics and engineering
        \item Physics and chemistry
    \end{itemize}\\
\hline

    Chemistry \& Biochemistry &
    \begin{itemize}
    \vspace{6pt}
        \item Chemistry and Biochemistry
        \item Biochemistry \& Cellular \& Molecular Biology
        \item Biochemistry
    \end{itemize}& 
     \begin{itemize}
        \item Biochem \& Molecular Biology
        \item Pharm Chem Sciences
        \item Chemical and Environmental Science 
    \end{itemize}\\
\hline

    Computer Science & 
    \begin{itemize}
        \vspace{6pt}
        \item Security Studies
        \item Cyber Operations and Resilience
        \item Systems engineering
        \item Informatics
        \item Information Systems
        \item Software and societal systems
        \item Data and Computational Science
        \item Data Science
        \item Applied Computation
        \item Computer and Information Science
        \item Geographic Information Systems
        \item Networking Technology
        \item Cyber Systems
    \end{itemize}
    & 
    \begin{itemize}
        \vspace{6pt}
        \item Security Studies
        \item Cyber Operations and Resilience
        \item Systems engineering
        \item Informatics
        \item Information Systems
        \item Software and societal systems
        \item Data and Computational Science
        \item Data Science
        \item Applied Computation
        \item Computer and Information Science
        \item Geographic Information Systems
        \item Networking Technology
        \item Cyber Systems
    \end{itemize}\\
\hline

    ECE &
    \begin{itemize}
            \vspace{6pt}
        \item Electrical and Computer Engineering
        \item Electrical Engineering
        \item Computer Engineering
        \item Computer Science and Engineering
        \item Electrical Engineering and Computer Science
        \item Computer Systems Engineering
        \item Electrical and Electronics engineering
        \item Electro-Mechanical Engineering Technology
        \item Computer engineering computer science
    \end{itemize} &
    \begin{itemize}
        \item Electrical Engineering \& Computer Science
        \item Electrical, Computer and Biomedical Engineering
        \item Electronics Engineering
        \item Electrical and Electronic Engineering
        \item Electrical, Computer, and Systems Engineering
        \item Intelligent Systems Engineering Department
        \item Electrical and Systems Engineering
        \item Electrical engineering, Mechanical engineering
    \end{itemize}\\
    \hline

    Mathematics &
    \begin{itemize}
                \vspace{6pt}
        \item Mathematics
        \item Mathematics - Discrete
        \item Committee on Computational and Applied Mathematics
    \end{itemize}&
    \begin{itemize}
        \item Applied Mathematics
         \item Applied and Computational Mathematics
        \item Mathematics, Science and Technology
    \end{itemize}\\ 
    \hline
\end{tabular}
\end{table*}

\begin{table*}[htb]
    \centering
    \begin{tabular}{|K{2.5cm}|m{7.5cm}m{7.5cm}|}
    \hline
    Materials Science (and Engineering) &
    \begin{itemize}
                \vspace{6pt}
        \item Materials Science and Engineering
        \item Material Engineering
        \item Material Science
        \item Materials Design and Innovation
        \item Materials Science, Engineering, and Commercialization
    \end{itemize}&
     \begin{itemize}
        \item Mechanical and Material Science
        \item Materials and Nanotechnology
        \item Science of Advanced Materials
        \item Materials and Biomaterials science and engineering
        \item Chemical and materials engineering
    \end{itemize}\\

\hline

    Other Engineering &
    \begin{itemize}
                \vspace{6pt}
        \item Microelectronic engineering
        \item Mechanical Engineering
        \item Bioengineering, Nanoengineering
        \item Mechanical and Aerospace Engineering, Nanoengineering
        \item Mechanical and Aerospace Engineering
        \item Nanoengineering, Chemical Engineering
        \item General Engineering
        \item Applied Science Engineering-Davis
        \item Mechanical \& Aeronautical Engineering
        \item Mechanical Engineering Technology
        \item Engineering and Applied Science
        \item Chemical \& Biomolecular Engineering
        \item Chemical Engineering
        \item Engineering Introduction
    \end{itemize}&
      \begin{itemize}
        \item Aerospace Engineering
        \item Interdisciplinary Engineering
        \item Metallurgical \& Materials Eng.
        \item Industrial Engineering
        \item Engineering and Environmental Science
        \item Biomedical engineering
        \item Aeronautics and Astronautics
        \item Mechatronics Engineering
        \item Photonics and Optical Engineering
        \item Nanoscience and Biomedical Engineering
        \item Engineering \& Technology
        \item Nuclear engineering
        \item Nuclear Engineering and Radiological Sciences
        \item Microsystems engineering
    \end{itemize}\\
\hline

    Miscellaneous STEM &
    \begin{itemize}
        \vspace{6pt}
        \item Biotechnology
        \item Science
        \item Nanoscale science
        \item Optical Sciences
        \item Applied Science and Technology
        \item NanoScience Technology Center
        \item Optics and photonics
        \item Radiological Sciences
        \item Biological sciences, Bioinformatics
        \item Environmental Science
        \item Cell and Molecular Biology
        \item Natural Sciences
        \item Geology
        \item Marine Geoscience
        \item Astronomy
        \item Electro-Optics
    \end{itemize}&
        \begin{itemize}
                    \vspace{6pt}
        \item Wond'ry Center for Innovation
        \item Biology
        \item Biosciences
        \item Vision Science
        \item Marine Sciences
        \item Energy Studies
        \item Academy of Integrated Science 
        \item Continuing Medical Education
        \item Earth Systems
        \item Technology Management
        \item Interdisciplinary Science
        \item Medical Imaging
        \item Nuclear Medicine Technology
        \item Certified Nurse Anesthesia
        \item Studies in Science, Technology and Faith
    \end{itemize}\\
\hline

    Non-STEM & 
    \begin{itemize}
         \vspace{6pt}
        \item Philosophy
        \item History 
        \item Doctor of Business Administration
        \item Logic and Philosophy of Science
        \item Political Science
        \item Creative Studies
        \item Music Therapy
        \item Curriculum and Instruction, School of Education
        \item Liberal Studies - Human Values
    \end{itemize}&
    \begin{itemize}
        
        \item Public Policy
        \item Communication 
        \item Religion
        \item Religious Studies
        \item History and philosophy of science
        \item Management
        \item Theology, Philosophy
        \item Public and Urban Policy
    \end{itemize}\\
\hline

    \end{tabular}
\end{table*}

\clearpage
\pagebreak
\section{Keywords for QISE Topics}
\label{keywords}
\begin{table}[htb]
    \centering
      \caption{ Keywords that were searched for in course descriptions. The left column keywords correspond to the QISE topics referenced throughout the paper. The right column lists all keywords included in the individual topics from the left column that were simplified for clarity in the figures. The lines spanning both columns are standalone for these purposes and appear in full in the body of the paper.}
    \begin{tabular}{|c|m{12cm}|}
    \hline
    \textbf{Topic discussed in paper} & \textbf{Individual keywords that makeup topic in searches.}\\
    \hline
    
    Gates & 
    \begin{itemize}
    \vspace{6pt}
    \item Gates
    \item CNOT
    \item Hadamard
    \item Pauli
    \item Swap
    \item Toffoli
    \end{itemize}\\
    \hline

    Algorithms & 
    \begin{itemize}
    \vspace{6pt}
        \item Algorithms
        \item Deutsch
        \item Grovers Algorithm
        \item Shors Algorithm
        \item Simons Problem
        \item Bernstein
    \end{itemize}\\
        \hline

    Cryptography & 
    \begin{itemize}
    \vspace{6pt}
        \item Cryptography
        \item Teleportation
        \item Key Distribution
        \item QKD
        \item Superdense
    \end{itemize}\\
    \hline

    Implementations &
    \begin{itemize}
    \vspace{6pt}
        \item Implementations
        \item Superconducting Qubits
        \item Trapped Ions
        \item Ion Trapping 
        \item Quantum Sensing
        \item Quantum Devices
        \item Quantum Hardware
    \end{itemize} \\
    \hline

    Advanced QISE Theory & 
    \begin{itemize}
    \vspace{6pt}
        \item NCT
        \item EPR
        \item Bells Inequality
        \item Decoherence
        \item Quantum Error
        \item Heisenberg Uncertainty
        \item Open Quantum Systems
    \end{itemize}\\
    \hline
    
        \multicolumn{2}{|c|}{Qubit}  \\
        \hline
    \multicolumn{2}{|c|}{Quantum Circuits} \\
        \hline
    \multicolumn{2}{|c|}{Quantum Computing}  \\
        \hline
    \multicolumn{2}{|c|}{Quantum Information}  \\
        \hline

\end{tabular}
  \label{tab:keywords}
\end{table}

\pagebreak
\section{Numerical values from figures in body of paper}
\label{chart_values}

\begin{table}[htb]
 \caption{Values corresponding to Fig.~\ref{fig:courses_by_discipline} \label{tab:courses_by_discipline}}
    \centering
    \begin{tabular}{|c|K{1.cm} K{1.cm} K{1.cm}|K{1.cm} K{1.cm} K{1.cm}|K{1.cm} K{1.cm} K{1.cm}|}
        \hline
        \multirow{2}{*}{\textbf{Discipline}} & \multicolumn{3}{c|}{\textbf{All Courses}} & \multicolumn{3}{c|}{\textbf{Courses with QISE Topics}} & \multicolumn{3}{c|}{\textbf{QISE Courses}} \\
        \cline{2-10}

         & \textbf{UG} & \textbf{DL} & \textbf{G} & \textbf{UG} & \textbf{DL} & \textbf{G} & \textbf{UG} & \textbf{DL} & \textbf{G} \\
         \hline
        \textbf{QISE-Specific} &  0 & 0 & 33 & 0 & 0 & 17 & 0 & 0 & 28\\
        \textbf{Physics} & 2874 & 312 & 1554 & 253 & 51 & 186 & 96 & 34 & 84 \\
        \textbf{Chemistry and Biochemistry} & 1316 & 151 & 517 & 18 & 1 & 18 & 0 & 0 & 0 \\
        \textbf{ECE} & 173 & 87 & 335 & 48 & 25 & 94 & 26 & 25 & 84 \\
        \textbf{Materials Science (and Engineering)} & 59 & 24 & 112& 5 & 0 & 12 & 0 & 0 & 4\\
        \textbf{Other Engineering} & 96 & 42 & 156 & 16 & 5 & 24 & 0 & 0 & 0 \\
        \textbf{Computer Science} & 101 & 30 & 128 & 87 & 27 & 98 & 49 & 22 & 57 \\
        \textbf{Mathematics} & 54 & 26 & 68 & 22 & 13 & 19 & 10 & 8 & 14 \\
        \textbf{Non-STEM} & 108 & 17 & 51 & 13 & 2 & 16 & 0 & 0 & 0 \\
        \hline
    \end{tabular}

\end{table}

\begin{table}[htb]
 \caption{Values corresponding to Fig.~\ref{fig:quantum_physics_course_types}}
    \centering
    \begin{tabular}{|c|K{1.75cm} K{1.75cm} K{1.75cm}|}
    \hline
        \multirow{2}{*}{\textbf{Course Type}} & \multicolumn{3}{c|}{\textbf{Physics Courses with `quantum'}}\\
                \cline{2-4}

        & \textbf{UG} & \textbf{DL} & \textbf{G} \\
        \hline
        \textbf{QISE} & 99 & 33 & 82 \\
        \textbf{QISE Adjacent} & 9 & 5 & 17 \\
        \textbf{Chemistry} & 17 & 0 & 6 \\
        \textbf{MISC} & 105 & 3 & 34 \\
        \textbf{Physics Electives} & 381 & 76 & 553 \\
        \textbf{Core Physics} & 2234 & 180 & 815 \\
        \hline
    \end{tabular}
   
    \label{tab:quantum_physics_course_types}
\end{table}

\begin{table}[htb]
    \caption{Values corresponding to Fig.~\ref{fig:quantum_physics_course_types}}
    \centering
    \begin{tabular}{|c|K{1.75cm} K{1.75cm} K{1.75cm}|}
    \hline
        \multirow{2}{*}{\textbf{Course Type}} & \multicolumn{3}{c|}{\textbf{Physics Courses with QISE Keywords}}\\
                \cline{2-4}

        & \textbf{UG} & \textbf{DL} & \textbf{G} \\
        \hline
        \textbf{QISE} & 90 & 32 & 65 \\
        \textbf{QISE Adjacent} & 1 & 0 & 0 \\
        \textbf{Chemistry} & 0 & 0 & 1 \\
        \textbf{MISC} & 13 & 0 & 4 \\
        \textbf{Physics Electives} & 20 & 3 & 32 \\
        \textbf{Core Physics} & 130 & 16 & 72 \\
        \hline
    \end{tabular}

    \label{tab:physics_courses_with_qise_topics}
\end{table}

\begin{table}[htb]
 \caption{Values corresponding to Fig.~\ref{fig:QISE_topic_prevalence}}
    \centering
    \begin{tabular}{|c|c|c|}
    \hline
        \textbf{QISE Topic} & \textbf{Non-QISE Frequency} & \textbf{QISE Frequency} \\
        \hline
        \textbf{Advanced QISE Theory} & 74 & 61 \\
        \textbf{Algorithms} & 28 & 241 \\
        \textbf{Circuits} & 4 & 33 \\
        \textbf{Cryptography} & 22 & 84 \\
        \textbf{Density Matrix} & 33 & 3 \\
        \textbf{Gates} & 73 & 35 \\
        \textbf{Implementations} & 26 & 46 \\
        \textbf{Quantum Computing} & 65 & 150 \\
        \textbf{Quantum Information} & 60 & 106 \\
        \textbf{Qubit} & 17 & 61 \\
        \hline
    \end{tabular}
   
    \label{tab:qise_topic_prevalence}
\end{table}

\clearpage
\twocolumngrid
\bibliography{bib} 

\end{document}